\documentclass[english,groupedaddress, superscriptaddress,pre,10pt,floatfix,reprint,twocolumn]{revtex4-1}
\usepackage[T1]{fontenc}
\usepackage[utf8]{inputenc}
\usepackage[normalem]{ulem}
\usepackage{amsmath,amsthm,mathtools,amssymb}
\usepackage{graphicx}
\DeclareMathOperator*{\argmax}{arg\,max}
\DeclareMathOperator*{\median}{Median}
\usepackage{longtable}
\usepackage{tabularx}
\usepackage{appendix}
\usepackage{listings}

\usepackage{units}

\usepackage{url}

\usepackage[breaklinks]{hyperref}

\usepackage[usenames,dvipsnames,svgnames,table]{xcolor}

\renewcommand{\d}{\mathrm{d}}

\renewcommand{\figurename}{Figure}
\renewcommand{\tablename}{Table}

\begin{document}

\title{Revealing drivers and risks for power grid frequency stability with explainable AI}

\author{Johannes Kruse}
\email{jo.kruse@fz-juelich.de}
\affiliation{Forschungszentrum J\"ulich, Institute for Energy and Climate Research - Systems Analysis and Technology Evaluation (IEK-STE), 52428 J\"ulich, Germany}
\affiliation{Institute for Theoretical Physics, University of Cologne, 50937 K\"oln, Germany}

\author{Benjamin Sch\"afer}
\email{b.schaefer@qmul.ac.uk}
\affiliation{School of Mathematical Sciences, Queen Mary University of London, London E1 4NS, United Kingdom}

\author{Dirk Witthaut}
\email{d.witthaut@fz-juelich.de}
\affiliation{Forschungszentrum J\"ulich, Institute for Energy and Climate Research - Systems Analysis and Technology Evaluation (IEK-STE), 52428 J\"ulich, Germany}
\affiliation{Institute for Theoretical Physics, University of Cologne, 50937 K\"oln, Germany}

\begin{abstract}
Stable operation of the electrical power system requires the power grid frequency to stay within strict operational limits. With millions of consumers and thousands of generators connected to a power grid, detailed human-build models can no longer capture the full dynamics of this complex system. Modern machine learning algorithms provide a powerful alternative for system modelling and prediction, but the intrinsic black-box character of many models impedes scientific insights and poses severe security risks. Here, we show how eXplainable AI (XAI) alleviates these problems by revealing critical dependencies and influences on the power grid frequency. We accurately predict frequency stability indicators (such as RoCoF and Nadir) for three major European synchronous areas and identify key features that determine the power grid stability. Load ramps, specific generation ramps but also prices and forecast errors are central to understand and stabilize the power grid. \end{abstract}

\maketitle

\section{Introduction}
The power grid frequency is the central observable in power system control, as it reflects the balance of power generation and demand \cite{machowskiPowerSystemDynamics2008}. An oversupply of power leads to an increase of the frequency, while a shortage causes a frequency drop. Large frequency deviations correspond to large power imbalances, which threaten system stability and may lead to large-scale blackouts \cite{pourbeikAnatomyPowerGrid2006}. Frequency stability is regarded as a major challenge in the transition to a sustainable energy system, as renewable power sources do not provide an intrinsic inertia \cite{milanoFoundationsChallengesLowInertia2018}. Hence, a better understanding of the emergence of large frequency deviations is urgently needed. 

Deviations from the reference frequency of 50/60 Hz arise due to distinct causes, modified by the complex interplay of different elements of the energy system. 
For example, the change of power generation at the intervals of electricity trading causes regular frequency jumps \cite{weissbachHighFrequencyDeviations2009}, whose magnitude depend on several technical parameters \cite{milanoFoundationsChallengesLowInertia2018,vorobevDeadbandsDroopInertia2019}. Fluctuating wind and solar power \cite{haehneFootprintAtmosphericTurbulence2018,ayodeleChallengesGridIntegration2012,collinsImpactsInterannualWind2018} or singular load patterns due to societal events \cite{chenAnalysisSocietalEvent2011}  create frequency fluctuations on different scales. To guarantee frequency stability in such a complex and uncertain environment, transmission system operators intensively monitor the system and allocate expensive control reserves. An improved understanding of the frequency dynamics and its interplay with external influences could greatly facilitate control efforts and contribute to power system stability.
	 
Modern methods from machine learning (ML) are excellent candidates for this task, with their capability of handling large amounts of features and data. In recent years, the amount of publicly available energy system data has grown steadily, including both frequency recordings \cite{gorjaoOpenAccessPowerGrid2020, krusePreProcessedPowerGrid2020a} and a variety of external features, such as generation and load time series  \cite{hirthENTSOETransparencyPlatform2018,morrisonEnergySystemModeling2018}. Thus, there is an optimal basis to analyze and predict the grid frequency with data-driven models \cite{hastieElementsStatisticalLearning2016}, even if interactions are non-linear and data are noisy. However, complex ML models are often non-explainable black-boxes, i.e. they do not provide insights about how inputs are mapped to outputs \cite{adadiPeekingBlackBoxSurvey2018a,roscherExplainableMachineLearning2020}. This is particularly problematic for critical infrastructures such as power systems, where the black-box character poses a security risk \cite{ahmadArtificialIntelligenceSustainable2021a, cremerOptimizationBasedMachineLearning2019}.

A solution are eXplainable Artificial Intelligence (XAI) approaches, a quickly growing research field, which covers inherently transparent ML models, as well as post-modelling explanations for black-box models \cite{barredoarrietaExplainableArtificialIntelligence2020}. SHapely Additive exPlanations (SHAP) values are such a post-modelling explanation, offering a unified way to measure feature effects and avoid inconsistencies present in other approaches \cite{lundbergUnifiedApproachInterpreting2017,lundbergConsistentIndividualizedFeature2019}. Only a few applications of this methodology have been presented in the broader field of energy systems analysis so far, for example to explain solar power forecasts \cite{kuzluGainingInsightSolar2020}, transient security assessments \cite{chenXGBoostBasedAlgorithmInterpretation2019} or power project failures \cite{alovaMachinelearningApproachPredicting2021}.

Here, we introduce an explainable ML model based on gradient boosted trees for selected indicators of frequency stability and evaluate its predictive power for three grids in Europe, the Continental Europe (CE), the Nordic and the Great Britain (GB) synchronous areas. We demonstrate the benefits of explainability via SHAPs,  going from coarse-grained global feature importances to detailed dependencies and finally to fine-grained interactions of different external features. Our approach complements established simulation based approaches that predict frequency deviations on the basis of load and generation forecasts. Simulations can be very accurate, but crucially depend on a faithful representation of the system and the quality of the input data and the forecasts. Data driven models can reveal additional driving factors, unknown effects and emerging risks and thus complement and improve simulations. For instance, our analysis highlights the role of forecasting errors, which varies strongly between the different grids under consideration.

\section{Results}

\subsection{Indicators of frequency stability}
\label{sec:indicators}

\begin{figure*}
	\centering
	\includegraphics[width=\textwidth]{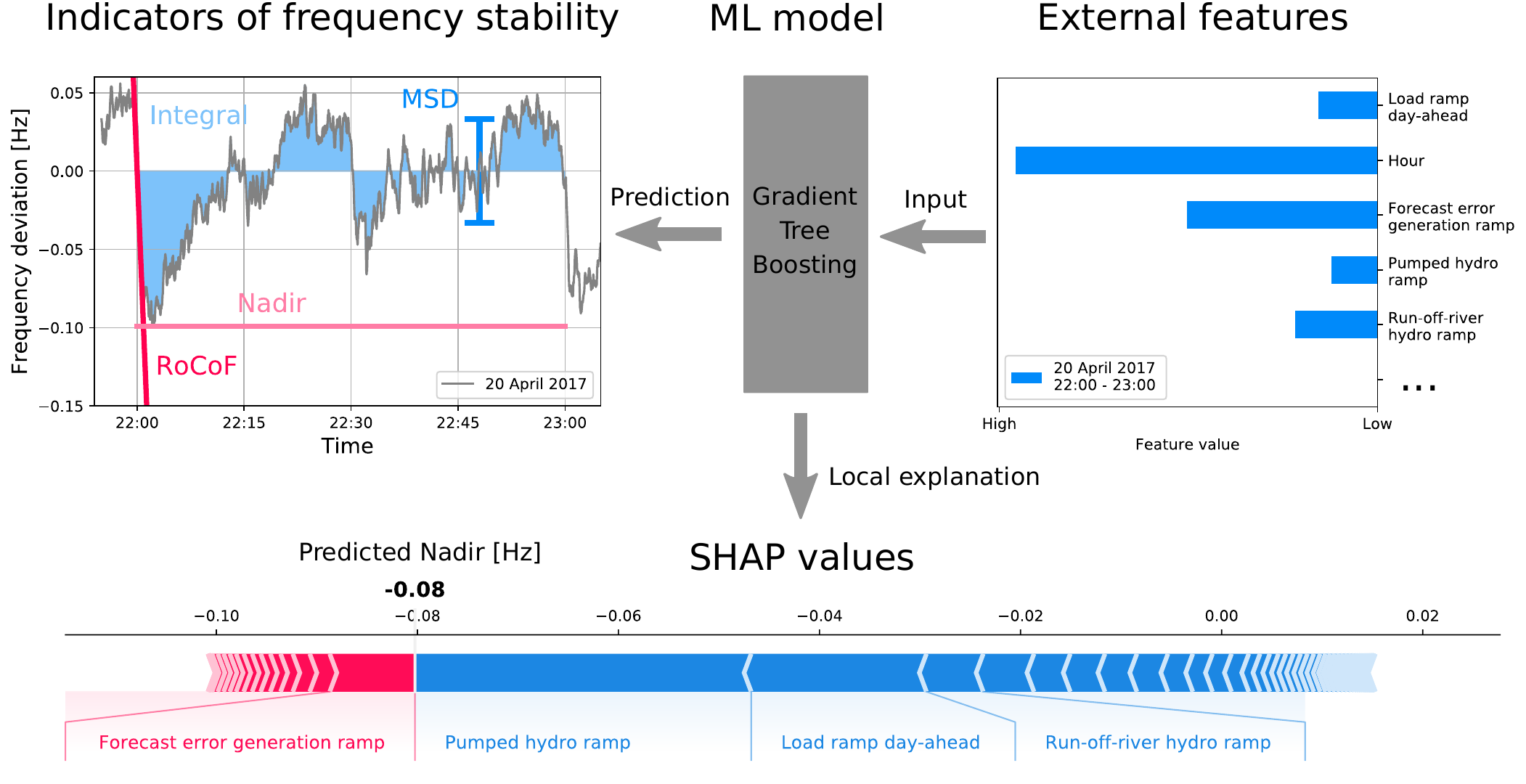}
	\caption{Sketch of our explainable ML model. From right to left: Using publicly available external features, such as load ramps or generation ramps \cite{ENTSOETransparencyPlatform}, we construct a Gradient Tree Boosting ML model that predicts indicators of frequency stability. The model is then interpreted via SHAP values, which quantify the effect of the features on the model output. The sketch depicts a sample hour of 2017 for Continental Europe.}
	\label{fig:1}
\end{figure*}

The power grid frequency fluctuates at various time scales ranging from seconds to weeks \cite{meyerIdentifyingCharacteristicTime2020}. In our model, we aggregate frequency deviations to hourly indicators that are directly relevant for power system stability (see Figure~\ref{fig:1} and Experimental Procedures). First, we analyze the maximum frequency deviation within the hour (Nadir) \cite{grossIncreasingResilienceLowinertia2017} and the Rate of Change of Frequency (RoCoF) \cite{grossIncreasingResilienceLowinertia2017}, which are of central relevance for grid monitoring and control. Nadirs above a threshold level indicate immediate danger and are counteracted by ultimate defence actions such as load shedding. Large RoCoFs are hazardous as control actions require a few seconds to take effect. In addition, we evaluate two integrated stability measures to account for the duration of frequency deviations. The variability of the hourly time series is characterized by the Mean Square Deviation (MSD) from 50 Hz. Notably, the MSD also indicates the total (primary) control effort, such that a large MSD reflects high operational costs \cite{tylooPrimaryControlEffort2021}. Finally, we evaluate the integrated frequency deviation (Integral), which is proportional to the mean deviation within the hour. Large integrals correspond to a systematic imbalance of the hourly power generation and the demand. 
We emphasize that regional differences of the grid frequency within a synchronous area remain small during normal operation and are typically damped out after several seconds
\cite{rydingorjaoOpenDatabaseAnalysis2020,gorjao2021phase}. Although we use local grid frequency measurements, the above  indicators thus characterize frequency stability globally in the whole synchronous area.

We evaluate these indicators on an hourly basis as this time scale is central for power system operation \cite{linElectricityMarketsTheories2017}. Electricity is traded predominantly in blocks of one hour and thus generation is adapted strongly at the beginning of the hour leading to deterministic patterns in the frequency \cite{weissbachHighFrequencyDeviations2009}. When the load decreases continuously during an hour, but the dispatch is set to the hourly mean, then power is scarce at the beginning of the hour and the frequency drops. As a consequence, frequency deviations show a pronounced daily profile, which we will later use as a null model to evaluate prediction performance. Moreover, most external features are only publicly available with hourly resolution \cite{hirthENTSOETransparencyPlatform2018}.

\subsection{An explainable model for frequency deviations}
\label{sec:model}

We introduce an explainable ML model that predicts indicators of frequency stability from external features (see Figure~\ref{fig:1} and Experimental Procedures). We use Gradient Tree Boosting, which produces non-linear models with state-of-the-art performance for many ML applications \cite{chenXGBoostScalableTree2016}, while enabling a fast and efficient calculation of SHAP values to explain the predictions \cite{lundbergLocalExplanationsGlobal2020a}. As inputs, we feed the model with physically meaningful features based on load, generation and electricity price time series. Our data includes both day-ahead available features, such as the day-ahead predicted load change ("Load ramp day-ahead"), and ex-post available features, such as the error between the day-ahead forecast and the actual total generation change ("Forecast error generation ramp"). Finally, we compute SHAP values to quantify how each feature contributes to the model output. For example, in Figure~\ref{fig:1}(bottom) the feature ``Load ramp day-ahead'' has a negative contribution (blue), thus causing the predicted Nadir to be lower than its average. SHAP values make local predictions more transparent, but they also enable aggregated insights about global feature effects, dependencies and interactions. 

Our model outperforms the daily average profile of the stability indicators, which is an important system-specific null model (see Experimental Procedures and Note S5). We achieve performances 3.2 (CE), 6.9 (Nordic) and 14.7 (GB) times higher than the daily profile, thus indicating additional important dependencies learnt by the model. Restricting the full model to day-ahead available features results in similar performance gains of 2.3 (CE) to 8.9 (GB), which opens the possibility of predicting stability indicators day-ahead. The additional benefit of including ex-post available features such as forecast errors is particularly large in the Nordic area. Here, the full model performs 2.5 times better than the restricted day-ahead model. This already points to the importance of forecast errors for the Nordic frequency dynamics, which we examine next.

\subsection{Main features affecting frequency deviations}
\label{sec:feature_importances}

\begin{figure*}
	\centering
	\includegraphics[width=\textwidth]{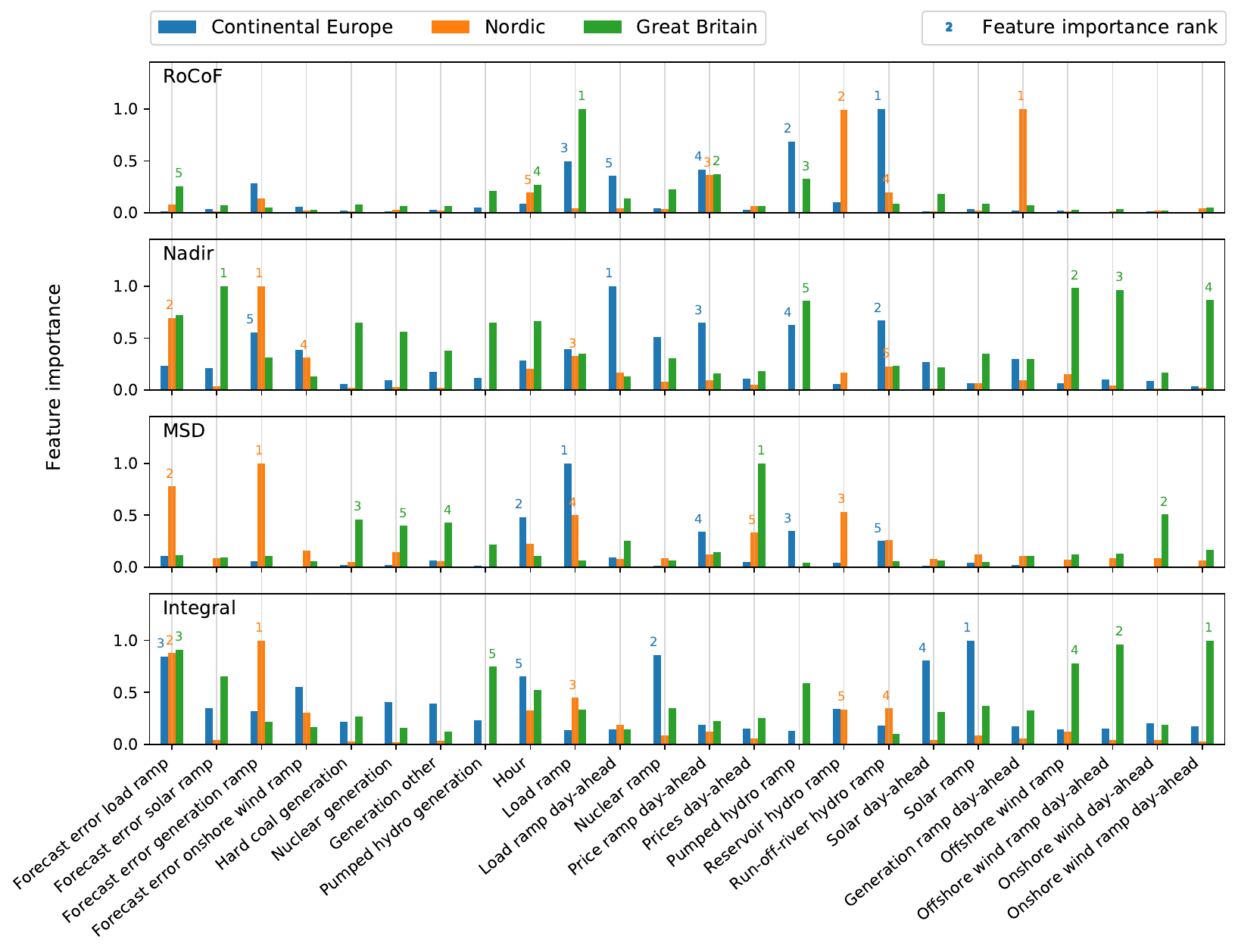}
	\caption{Most important features for predicting frequency stability. The feature importance in our model is measured by the mean absolute SHAP value (see Experimental Procedures). We highlight the five most important features by displaying their importance rank above the corresponding bar. While forecast errors, load and generation ramps have a high relative importance, the total synchronous generation does not appear among the important features and its average effect is therefore negligible.}
	\label{fig:2}
\end{figure*}

We start demonstrating our model explainability on the coarse-grained level of global feature importances, which characterize how much a certain feature affects the hourly frequency stability indicators within the trained model (Figure~\ref{fig:2}). 

In the RoCoF model, only a few features dominate, which are mainly generation ramps from hydropower and load ramps. The importance of hydro ramps relates to their large ramping speed, which we discuss below.  In the Nordic area, the total day-ahead generation ramp is much more important for the RoCoF than load ramps. This suggests that changes of power export and storage may be relevant here, as these are not represented in the load of the area.

The Nadir is primarily affected by ramps and their respective forecasting errors. In CE, the load ramp is the most important feature by a large margin. This reflects the importance of deterministic frequency jumps, which are strongly correlated with the direction of the load ramps \cite{weissbachHighFrequencyDeviations2009}. In the Nordic grid, the forecast errors of generation and load ramp are by far the most important features, partly explaining the large performance gain when including ex-post data in the model (see above). In contrast, there are many almost equally important features in the British Nadir model, with wind power generation and ramps and solar ramp forecast errors peaking out. This indicates the importance of renewables for frequency deviations in GB.

The MSD behaves similarly to the Nadir in CE and the Nordic grid, with some subtle differences. The load ramp is the most important feature in CE. Forecasting errors again dominate in the Nordic grid, but reservoir hydro ramps now also contribute. A different situation is found in GB. Absolute variables such as the day-ahead prices and wind power generation dominate the MSD prediction, while generation ramps do not significantly contribute. These differences point to a more complex behaviour of the MSD, which we will further investigate below.

Finally, the Integral is mainly explained by forecasting errors of load and generation ramps, as such errors cause long-lasting power mismatches. This is particularly evident in the Nordic grid, where other features have significantly lower importance. In GB, wind power ramps are ranked highly, confirming the importance of renewables. In CE, solar and also nuclear power ramps are relevant for the prediction. We will investigate how the interaction of these two distinct generation types explains model variance below.

In summary, CE exhibits strong deterministic frequency deviations (DFDs) that are connected to hourly load and generation ramps (cf.~\cite{weissbachHighFrequencyDeviations2009, entso-eaisblReportDeterministicFrequency2019}). Nordic frequency deviations are mainly connected to forecasting errors and reservoir hydro ramps (cf.~\cite{svenskakraftnatChallengesOpportunitiesNordic2016}). In GB, hourly DFDs are less important (see also Note S4) and frequency deviations are mainly affected by renewables, i.e. their fluctuations and forecast errors (cf.~\cite{nationalgridesoOperabilityStrategyReport2019}). The total synchronous generation, which approximates the inertia in our model (see Experimental Procedures), affects the British frequency dynamics only in extreme situations with very low inertia (Note S6). 
In spite of the importance of reduced inertia in renewable energy systems \cite{milanoFoundationsChallengesLowInertia2018,nationalgridesoOperabilityStrategyReport2019},  our model shows that the average effect of inertia on the aggregated stability indicators is negligible (Figure~\ref{fig:2}).  This is consistent with other studies on aggregated frequency fluctuations (cf. \cite{vorobevDeadbandsDroopInertia2019}), which found that inertia is important for extreme events but not the aggregated dynamics.

\subsection{Characterizing the effect of generation ramps}
\label{sec:RoCoF_effects}

\begin{figure*}
	\centering
	\includegraphics[width=\textwidth]{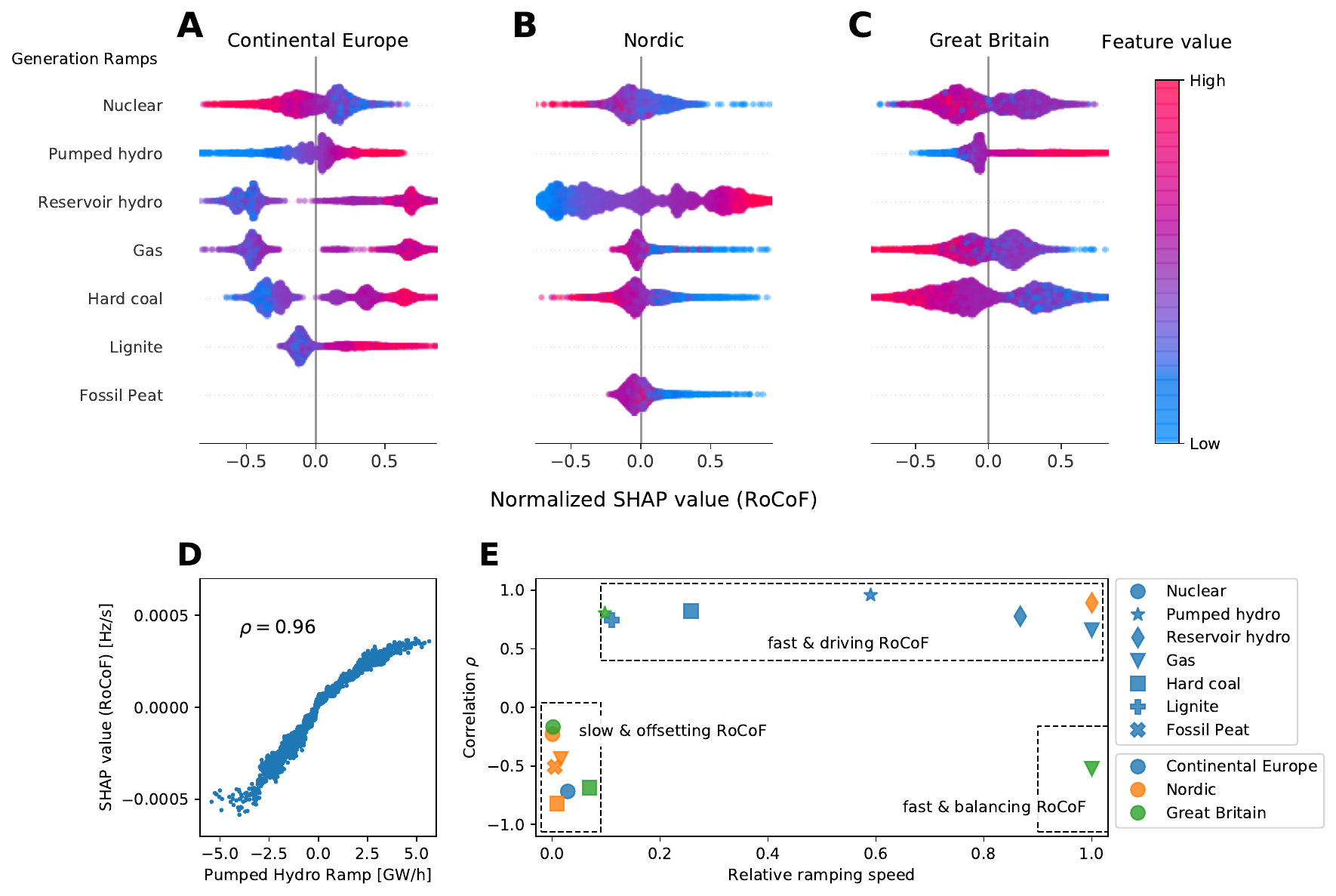}
	\caption{Effect of generation ramps on the RoCoF. (A-C): We examine the effects of dispatchable, i.e. weather-independent generation technologies, which mostly affect the hourly RoCoF due to their step-wise change at the beginning of (hourly) trading intervals \cite{weissbachHighFrequencyDeviations2009}. The summary plots depict the SHAP effects on the RoCoF normalized by their maximum, absolute value to improve visibility. The figure only examines generation ramps with a feature importance above 0.02 to ensure the reliability of the results. (D): We quantify the direction of the dependencies with the correlation $\rho$ between the feature value and the SHAP effect, as demonstrated here for pumped hydro ramps. (E): Combining the directions $\rho$ with the relative ramping speeds of the generation technologies (see Experimental Procedures) enables us to distinguish RoCoF-driving and RoCoF-offsetting technologies within the areas.}
	\label{fig:3}
\end{figure*}

Fast generation ramps significantly affect the hourly RoCoF. Going beyond mean feature importances, we now examine the direction of these dependencies via SHAP values (Figure~\ref{fig:3}). The effect of ramps is mostly monotonic, that is the feature effect increases or decreases monotonically with the feature value (Figure~\ref{fig:3}A-C). Remarkably, the direction of the dependency varies among the different generation types and grids. For instance, the dependency of hard coal ramps in CE is opposite to the behaviour in GB and the Nordic grid.

We explain the observed differences in terms of the \emph{relative} ramping speed of a generation type within the respective area (see Experimental Procedures). In the Nordic grid, hydropower is essential and all other generation types must be considered slow in comparison. In GB and CE, non-hydropower dominates the generation mix and technologies with slower ramps than hydropower plants but ramps faster than other generation types play important roles. Notably, hard coal belongs to the slow generation types in GB but to the fast types in CE due to the importance of nuclear and lignite generation in CE, which are even slower. Now, we utilize SHAP values of generation ramps when predicting the RoCoF and relative ramping speed to categorize generation types (Figure~\ref{fig:3}D-E). We find that fast generation ramps drive the RoCoF: A positive ramp is associated with more positive frequency jumps. In contrast, ramps of slow generation types offset the RoCoF, leading to a negative correlation. The only exception is the behaviour of gas power plants in GB, which show a negative correlation despite being fast. This can be explained by their role as the prime source of balancing reserve in GB \cite{nationalgridesoMonthlyBalancingServices2019}. Loosely speaking, the ramps no longer drive the RoCoF, but the RoCoF drives the ramps. 

Notably, a model-agnostic data analysis cannot reveal such consistent results, as our features are strongly correlated (see Experimental Procedures). For example, the Pearson correlation coefficient between nuclear ramps and RoCoFs in CE is positive (Note S3). Instead, the SHAP framework indicates a negative relationship, which we consistently explain with relative ramping speeds.

\subsection{Relating large control efforts to non-linear dependencies}
\label{sec:control_effort}

\begin{figure*}
	\centering
	\includegraphics[width=\textwidth]{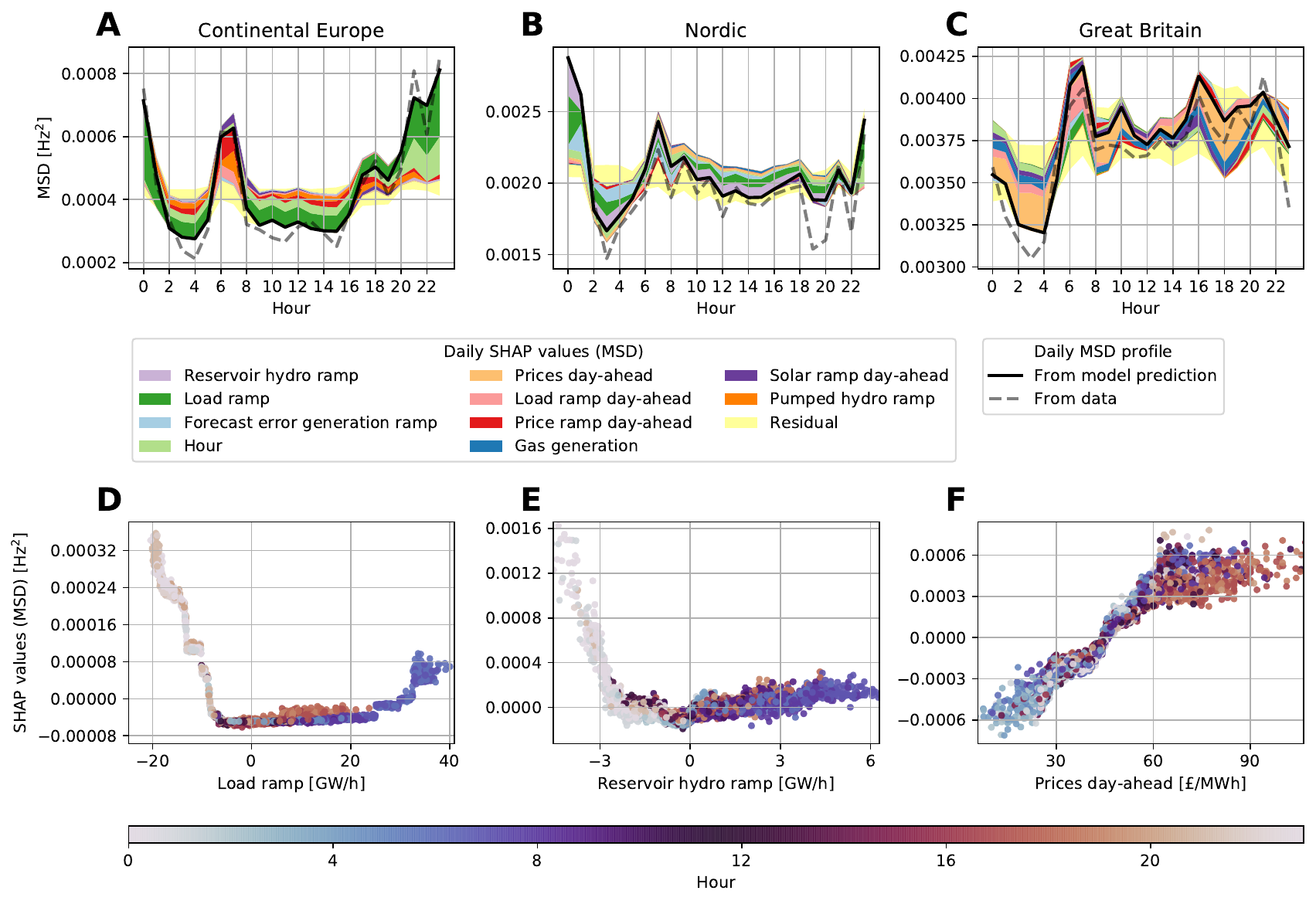}
	\caption{Explaining the daily average control effort with SHAP values. (A-C): The daily average profile of the MSD (dashed line), i.e. the daily average control effort, is very well reproduced by the ML model (solid line), but its shape differs among the areas. We explain the differences with daily SHAP values (see Experimental Procedures), which are sorted such that negative effects are plotted above the prediction line and positive effects below. The importance of the plotted feature effect decreases with growing distance to the prediction line and less important features are aggregated in a residual variable. (D-F): For the Nordic and the CE areas, the dependency plots of the most important daily SHAP effects reveal non-linear relationships. These explain the large control effort around midnight (colour code), while the linear effect of prices explains the low control effort in GB during the night.}
	\label{fig:4}
\end{figure*}

Frequency stability indicators often exhibit a complex non-linear dependency on the features. Using the MSD, an indicator for the (primary) control effort \cite{tylooPrimaryControlEffort2021}, as an example, we first note that the daily profiles of the  MSD starkly differ between the three grids (Figure~\ref{fig:4}). These differences are well reproduced by the ML model and we explain them via daily aggregated SHAP values (see Experimental Procedures). 

In CE, the control effort peaks around midnight (Figure~\ref{fig:4}A), which is strongly related to the non-linear effects of negative load ramps. Details about this relation are revealed by a dependency plot (Figure~\ref{fig:4}D). Load ramps between -7 GW/h and +25 GW/h have a small negative effect on the MSD, relating to the fact that such small ramps are easy to control. Outside this interval, the effect grows strongly in a non-linear and asymmetric way.  Negative load ramps have much larger effects than positive ones, and occur almost exclusively around midnight (see colour code). The most important feature in the Nordic daily profile are reservoir hydro ramps (Figure~\ref{fig:4}B), also showing a pronounced nonlinear dependency (Figure~\ref{fig:4}E). Large negative ramps have a much stronger effect than large positive ramps and explain a large part of the MSD peak around midnight. In contrast, the daily MSD profile in GB strongly depends on day-ahead prices (Figure~\ref{fig:4}C), which display an almost linear dependency (Figure~\ref{fig:4}F). The control effort peaks during the day, which is strongly connected to high prices in the day-ahead market, while the MSD and the prices are low during the night (00:00 to 04:00). 

Notably, fluctuating renewables do not contribute strongly to the daily MSD profile in our model, although they are an important driver for frequency fluctuations in GB in general revealed by their feature importance. The observed stark differences between the synchronous areas might be explained by different control regulations. For example, in GB wind power farms have to provide frequency control \cite{diaz-gonzalezParticipationWindPower2014a} and secondary control is allocated manually \cite{entso-eENTSOEBalancingReport}.

\subsection{Explaining systematic imbalances with interactions}
\label{sec:interactions}

\begin{figure*}
	\centering
	\includegraphics[width=\textwidth]{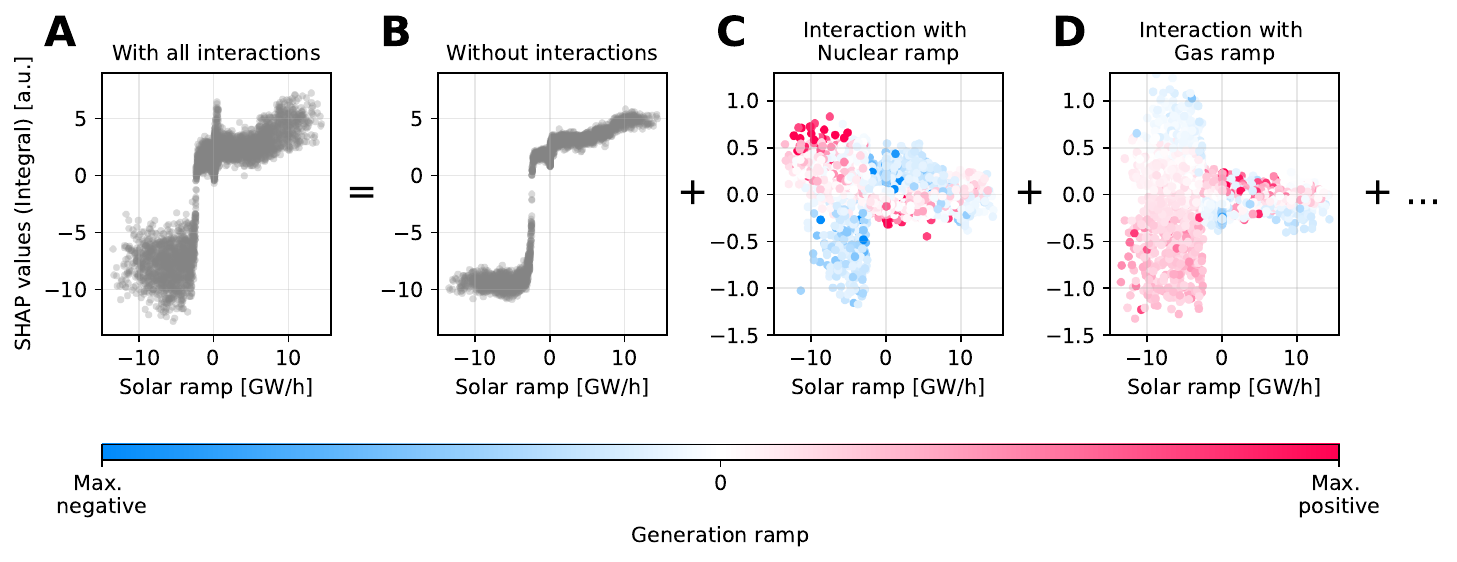}
	\caption{The effect of feature interactions on systematic power imbalances. (A): Taking CE as an example, we show the SHAP effects of solar ramps on the frequency integral, which are the most important effects in CE (Figure~\ref{fig:2}). The integral, which represents systematic imbalances, decreases for negative solar ramps, but the SHAP effects can vary strongly as indicated by their vertical dispersion. (B-D): Using SHAP interaction values (see Experimental Procedures), we decompose this dispersion into different interaction effects. These depend on the generation type, as negative nuclear ramps weaken the effect of negative solar ramps, while negative gas ramps relate to an amplified effect.}
	\label{fig:5}
\end{figure*}

The SHAP framework allows disentangling the role of different features and even reveals how the prediction depends on the \emph{interaction} of features (see Experimental Procedures). We demonstrate this for the Integral in the CE grid (Figure~\ref{fig:5}). The most important features are solar and nuclear power ramps, however with a reverse dependency (Figure S13). Without interactions, the SHAP value increases non-linearly with the solar ramp in an almost step-like fashion (Figure~\ref{fig:5}B). Strong negative ramps of solar power generation induce an ongoing shortage of power and thus lead to negative Integrals.

Interactions with nuclear and gas ramps can alter the effect of solar ramps by up to 50\%, leading to a strong vertical dispersion of the observed SHAP values (Figure~\ref{fig:5}A). In particular, negative nuclear ramps amplify the effect of negative solar ramps, while negative gas ramps dampen the effect on the Integral (Figure~\ref{fig:5}C-D). These opposite interactions can again be related to different ramping speeds. Nuclear power has the lowest ramping speed in CE, such that negative nuclear ramps can amplify the continuous ramping behaviour in interaction with solar ramps. In contrast, gas power is ramping fast and therefore often provides balancing power, leading to an opposite effect. In general, this demonstrates that interactions can severely influence how strongly a single feature affects power system stability. 

\section{Discussion}
\label{sec:discussion}

In summary, we have introduced an explainable machine learning model to predict important indicators for power system frequency stability from external features such as day-ahead electricity prices or the total system load. Using actual data, the ML model outperforms the daily profile, a system-specific null model, by a factor of up to 14.7. Using only day-ahead available data, the ML models show similar performance in most cases, paving the way for immediate applications. A notable exception is the Nordic grid, pointing to a strong role of forecasting errors in this synchronous grid.

Our model has been explained via SHAP values, which offer a consistent way to analyze the dependencies of features and predictions. This framework has confirmed the importance of forecasting errors for Nordic frequency deviations and allowed us to characterize the effect of generation ramps on the hourly Rate of Change of Frequency (RoCoF), which explains the different area-specific roles of generation technologies for deterministic frequency deviations. Furthermore, our dependency analysis of the primary control effort quantified by the MSD has pointed to unexpected non-linear relations and differences between the grid areas, which have to be further explained. Finally, we have demonstrated how SHAP interaction values distinguish the role of different conventional generation technologies in power grids with renewable power sources. 

The main restrictions to our model performance and explainability arise due to the quality of available power system data. First, frequency deviations due to renewable fluctuations \cite{haehneFootprintAtmosphericTurbulence2018} or load fluctuations \cite{chenAnalysisSocietalEvent2011} occur on time scales that are smaller than the intervals of electricity trading. The limited time resolution of publicly available power system time series restricts both the performance of an ML model and its ability to suggest causal relationships since the time order of events is partly hidden. Secondly, all locations in a synchronous power grid affect the frequency deviations, but in large grid areas, such as Continental Europe, many countries provide no or only a limited amount of data \cite{hirthENTSOETransparencyPlatform2018}. This further emphasizes the need for open data in energy system analysis and design \cite{pfenninger2017energy}. 

In conclusion, we hope that our work triggers further applications of explainable AI in energy science, harnessing the strengths of modern machine learning tools while avoiding the drawbacks of black-box approaches, which impede scientific insights \cite{roscherExplainableMachineLearning2020} and pose security risks \cite{ahmadArtificialIntelligenceSustainable2021a}. With our model, we gain insights by explaining feature effects with SHAP values in the context of the domain science. SHAP dependency and interaction plots visualize the knowledge learnt by the model and offer individual explanations for each prediction.  The most predictive associations then suggest causal relationships, which have to be validated by domain knowledge or further experiments. For example, we identify RoCoF-driving, RoCoF-offsetting and RoCoF-balancing generation technologies by connecting our model results to physical ramping rates, thus suggesting different causal relationships. Furthermore, our model points to unexpected dependencies, reveals differences among the grid areas, and suggests measures to improve frequency stability, for instance via improving forecasts in the Nordic grid. All in all, SHAP values alone do not give scientific insights, but together with domain knowledge, they can lead to discoveries.

\section{Experimental procedures}
\label{sec:methods}
\subsection{Resource Availability}
\subsubsection*{Lead contact}
Further information questions should be directed to the lead contact, Johannes Kruse (jo.kruse@fz-juelich.de).
\subsubsection*{Materials availability}
This study did not generate new unique materials.
\subsubsection*{Data and Code Availability}
The data set to reproduce our results is available on zenodo \cite{kruseDataSetPredictionFrequency}. The python code used to create our results and the figures is available on GitHub \cite{frequency_xai_repo}. 

\subsection{Data preparation of frequency stability indicators}
In a modern AC power grid, the grid frequency is typically spatially synchronized and its dynamics can be represented by a single bulk time series on time scales of several seconds and larger \cite{machowskiPowerSystemDynamics2008}. In Europe, different synchronous areas exist, which are only inter-connected with DC-links and hence display their own frequency dynamics and follow their specific regulations. We model the bulk frequency dynamics for different synchronous areas in Europe, particularly for the Continental Europe (CE), the Nordic and the Great Britain (GB) areas. We use pre-processed frequency time series $\tilde f(t)$ with a time resolution of $\tau=1$s  \cite{krusePreProcessedPowerGrid2020a}, which were originally measured by regional transmission system operators \cite{RegelenergieBedarfAbruf, HistoricFrequencyData,FrequencyHistoricalData}. 

From the centered frequency time series $f(t) = \tilde f(t) - 50$ Hz, we extract four hourly stability indicator, which are directly relevant for power system operation \cite{grossIncreasingResilienceLowinertia2017, tylooPrimaryControlEffort2021}. For the $i$-th hour starting at time $t_i$, we calculate the (positive or negative) Nadir, the Integral and the MSD based on the hourly time steps $\mathcal I_i = \{t_i, t_i + \tau, ..., t_i + \tau\gamma\}$ with $\gamma =3600$:

\begin{align*}
    \textrm{Nadir}(t_i) &= f\left(\argmax_{t\in \mathcal I_i} \left| f(t) \right| \right), \\
    \textrm{Integral}(t_i) &= \tau \sum_{t \in \mathcal I_i} f(t), \\
    \textrm{MSD}(t_i) &= \frac{1}{\gamma}\sum_{t \in \mathcal I_i} f^2(t)
\end{align*}

From the derivative of the frequency time series $\frac{\d f}{\d t}(t)$, we obtain the hourly (positive or negative) RoCoF by looking for the steepest slope within a window $\mathcal W_i = [t_i-T,t_i+T]$ of length $2T$ around the beginning of the hour $t_i$:

\begin{align*}
    \textrm{RoCoF}(t_i) = \frac{\d f}{\d t}\left( \argmax_{t \in \mathcal W_i } \left|\frac{\d f}{\d t}\right| \right)
\end{align*}

We estimate the derivative $\frac{\d f}{\d t}(t)$ by using a low-pass filter on the frequency increments \cite{frigoPMUBasedROCOFMeasurements2019}, i.e. by smoothing the increments $\Delta f(t) = f(t)-f(t-\tau)$ with a rectangular rolling window of length $L$. We choose the parameters $L$ and $T$ in such a way that they account for the different time scales of the RoCoF in the synchronous areas (Note S2). This results in a choice of $L=T=60$s for the CE and GB areas, while the Nordic area with its large amount of fast hydropower exhibits larger Rocofs so that we choose $L=T=30$s instead.

\subsection{Data preparation of external features}
We collect different power system time series as external features to predict frequency deviations in Europe. We retrieve six different sets of publicly available time series from the ENTSO-E Transparency platform \cite{ENTSOETransparencyPlatform}. These sets comprise the day-ahead load forecast, day-ahead scheduled generation, day-ahead wind and solar power forecast, day-ahead electricity prices, actual load and actual generation per production type. Most of the time series are available on an hourly basis. Since we predict stability indicators on an hourly basis, we down-sample a few higher resolution time series to a common hourly resolution. 

Then, we aggregate the data within the three synchronous areas. Since time series from ENTSO-E are only available for smaller regions within the synchronous areas (e.g. countries), we sum up the load and generation data within each area. To aggregate the price data, we calculate area-wide averages weighted by the regional mean load. The time series from the ENTSO-E Transparency platform contain multiple missing or corrupted data points \cite{hirthENTSOETransparencyPlatform2018}, which requires a careful aggregation and cleansing procedure (Note S1). All locations within the synchronous power grid contribute to large frequency deviations \cite{machowskiPowerSystemDynamics2008}, which motivates the area-wide feature aggregation. To support this, we additionally prepare selected country-level data for the CE and the Nordic areas. The area-wide aggregated features result in a similar or higher model performance than country-level data (Note S5). Therefore, we use area-wide aggregated features within this publication. An overview of the available (aggregated) features per area is available in Note S1. 

Finally, we engineer additional meaningful features based on the hourly ENTSO-E time series $X(t_i)$, which comprise both day-ahead forecast data $X_{D-1}(t_i)$ and actual data $X_{D}(t_i)$. For each hourly interval $\Delta t = \tau\gamma$, we introduce ramp features (i.e. slopes) $\left(X(t_i)-X(t_i-\Delta t)\right)/\Delta t$, which are inspired by the importance of generation ramps for the CE frequency dynamics \cite{weissbachHighFrequencyDeviations2009}. We further add forecast errors $X_{D-1}(t_i)- X_{D}(t_i)$ and the artificial features hour (of the day), weekday and month. To include the total available inertia as a feature, we calculate the sum of the synchronous generation which approximates the time-dependent inertia \cite{ulbigImpactLowRotational2014}. 

In summary, our data set comprises hourly time series of four stability indicators (model outputs or targets) and 66 external features (model inputs) for the years 2015- 2019. The data set is available on zenodo \cite{kruseDataSetPredictionFrequency} and our scripts for downloading and preparing the data set are online at GitHub \cite{frequency_xai_repo}.

\subsection{Gradient tree boosting model}
To predict indicators of frequency stability from external features, we use Gradient Tree Boosting (GTB). Tree-based ensemble methods such as GTB are complex, non-linear machine learning models, which makes them suitable to predict the non-linear behaviour of power grids \cite{machowskiPowerSystemDynamics2008}. They further offer a fast way to calculate SHAP values thus enabling an efficient post-modelling explanation \cite{lundbergLocalExplanationsGlobal2020a}. Additionally, they are immune to the effects of feature outliers and perform inherent feature selection thus being robust to the inclusion of correlated or irrelevant features \cite{hastieElementsStatisticalLearning2016}. This is beneficial for our data set, which exhibits strongly correlated features (Note S3) as well as outliers due to bad data quality (Note S1). 

To fit a GTB model, we use XGBoost, a scalable gradient tree boosting system that gives state-of-the-art results for many ML applications \cite{chenXGBoostScalableTree2016}. We randomly split our data into a training set (64\%), a validation set (16\%) and a test set (20\%). To optimize the hyper-parameters of the XGBoost model, we perform a grid search over selected parameter values and evaluate the performance via 5-fold cross-validation on our training set. To determine the number of trees in the XGBoost models, we perform early stopping on the validation set. Finally, we train the model with optimal hyper-parameters on the training set and test its performance on the unseen test set. The detailed implementation, in python code, is available on GitHub \cite{frequency_xai_repo} and the sets of final hyperparameters are online at zenodo \cite{kruseDataSetPredictionFrequency}. 

To quantify the model performance, we evaluate the R$^2$-score, which quantifies the proportion of variability explained by the model. Predicting the true targets results in a score of 1, while always predicting the mean of the target gives a score of 0. To benchmark our predictor, we compare its performance to the daily profile prediction. The daily profile, i.e. the daily average evolution of a target, is the most important recurring pattern of frequency dynamics \cite{krusePredictabilityPowerGrid2020}. Predicting the stability indicators based on their daily profiles thus represents an important null model. Our GTB model always outperforms the daily profile for all areas and indicators, see Note S5 for a detailed performance evaluation.

\subsection{Model interpretation with SHAP}
SHapely Additive eXplanation (SHAP) values offer a unified way to explain the output of any machine learning model \cite{lundbergUnifiedApproachInterpreting2017}. Based on the game-theoretical Shapely values, they attribute a model output to the individual effects of each input feature. Within the class of additive feature attributions, they uniquely guarantee certain optimal properties such as local accuracy and consistency, thus being in line with human intuition \cite{lundbergConsistentIndividualizedFeature2019}. Due to local accuracy, the SHAP values always add up to the total model output. Consistency guarantees, that a SHAP value does not decrease if the corresponding feature contributes more to the prediction due to a change of the model. 

SHAP values quantify feature effects on individual model outputs relative to the average prediction (the base value $\phi_0$). By combining many of these local explanations, SHAP values also offer global insights \cite{lundbergLocalExplanationsGlobal2020a}. The mean absolute SHAP value measures the global importance of a feature within the model. To give an overview of the important features in Figure~\ref{fig:2}, we collect the five most important features for each stability indicator and area. The figure displays feature importances for the \textit{union} of these feature sets, i.e. also features with an importance rank below five are displayed. In addition to global feature importances, dependency plots show how the effect of a feature changes with the value of the feature, such as in Figure~\ref{fig:4}D. Notably, these dependencies differ from observing relationships in scatter plots or in a simple correlation analysis between targets and features. Such model-agnostic methods suffer from feature correlations, i.e. they cannot distinguish the effect of two correlated features. In contrast, we estimate interventional SHAP values, which quantify the marginal feature effect in the model by breaking the correlations with other features \cite{janzingFeatureRelevanceQuantification2020,chenTrueModelTrue2020}. 

In addition to first-order attributions, SHAP offers interaction values that attribute the model output to pairs of interacting features \cite{lundbergLocalExplanationsGlobal2020a}. Interaction values decompose the first-order SHAP effects into diagonal effects and pairwise interaction effects (such as in Figure~\ref{fig:5}). In that way, the interaction effects explain the vertical dispersion within the first-order SHAP dependency plots thus offering scientific insights as well as additional consistency checks for the model applications. 

To explain daily average profiles of the model predictions, we introduce a new application for SHAP values.
The daily profile of the prediction $f(t_i)$ is the average $\langle f(t) \rangle_h$ for the hour $h$ over all days. Based on the SHAP values $\phi_j(t_i)$ for feature $j$ ($j=1,...,N$) and their base value $\phi_0$  \cite{lundbergLocalExplanationsGlobal2020a}, we decompose the daily profile as follows:

\begin{align*}
    \langle f(t) \rangle_h = \langle \phi_0 + \sum_{j=1}^{N} \phi_j(t_i)\rangle_h = \phi_0 + \sum_{j=1}^{N} \langle \phi_j(t) \rangle_h .
\end{align*}

The daily aggregated SHAP values $\langle \phi_j \rangle_h$ then explain the daily profile of the prediction. To display the daily SHAP values, such as in Figure~\ref{fig:4}A-C, we collect the four most important features according to their average effect $\frac{1}{24}\sum_{h=1}^{24} | \langle \phi_j \rangle_h|$ on the daily profile in each area. In Figure~\ref{fig:4}A-C, we then visualize the features from the union of these sets to display the most important daily SHAP values. The remaining daily SHAP values are aggregated and displayed as a residual variable.  

\subsection{Relative ramping rates}
We use relative ramping rates to validate our SHAP results, especially for the prediction of the RoCoF. Therefore, we quantify the relative ramping speed of each conventional generation technology $k$ within a synchronous area. The ramping speed $\tilde s_k$ is determined both by the absolute change of generation $\Delta X_k$ and the time scale $\lambda_k$ on which the generator adapts its output to the new set point:

\begin{align*}
\tilde s_k:=\frac{\Delta X_k}{\lambda_k}.
\end{align*}

We approximate the typical value of $\Delta X_k$ with the median of the absolute generation changes $\Delta X_k \approx \median_{t_i}  |X_k(t_i-\Delta t) - X_k(t_i)|$. The \textit{relative} ramp speed $s_k$, as compared to the fastest technology $m$ within the area, then reads

\begin{align*}
    s_k = \frac{\tilde s_k}{\tilde s_m} = \frac{\Delta X_k}{\Delta X_{m}} \frac{\lambda_{m}}{\lambda_k}\approx \frac{\Delta X_k}{\Delta X_{m}} \frac{r_k}{r_m} .
\end{align*}

In the last step, we approximate the ratio of time scales $\lambda_k$ with the inverse ratio of technology-specific ramping rates $r_k$ \cite{kondziellaFlexibilityRequirementsRenewable2016}. The technology $m$ with the largest absolute ramping speed is determined by the maximum value of $\Delta X_k r_k$. 

\section{Supplemental information}

Document S1. Notes S1-S6, Figures S1–S14, and Tables S1-S2.

\section{Acknowledgements}

We thank Bo Tranberg for fruitful discussions. Furthermore, we gratefully acknowledge support from the German Federal Ministry of Education and Research (BMBF grant no. 03EK3055B) and the Helmholtz Association via the \textit{Helmholtz School for Data Science in Life, Earth and Energy} (HDS-LEE). This project has received funding from the European Union’s Horizon 2020 research and innovation programme under the Marie Sk\l{}odowska-Curie grant agreement No. 840825.

\section{Author contributions}
J.K., B.S., D.W. conceived and designed the research. J.K. trained the model and produced the figures. All authors contributed to discussing and interpreting the results and writing the manuscript. B.S. and D.W. contributed equally.

\section{Competing interests}
The authors declare no competing interests.

\bibliographystyle{naturemag}
\bibliography{references}

\clearpage

\section*{Supplemental information} 
\setcounter{figure}{0}
\renewcommand{\thefigure}{S\arabic{figure}}
\renewcommand{\thetable}{S\arabic{table}}
\renewcommand{\figurename}{Figure}
\renewcommand{\tablename}{Table}

\section*{Note S1: External feature aggregation and data cleansing}
\label{sec:supp_note_feature_details}
To model frequency stability indicators, we collect publicly available times series of external features from the ENTSO-E Transparency platform \cite{ENTSOETransparencyPlatform}. For the synchronous areas investigated here, we aggregate the ENTSO-E time series, which are originally only available for smaller regions within the areas (e.g. countries). However, the time series contain many missing data points, so that we need a careful procedure to aggregate the region contributions within the synchronous areas.

Firstly, we specify region types for which we obtain the best data quality. In all but a few cases, we retrieve country level data. Only in Continental Europe, we retrieve bidding zone data for Italy (North, Center North, Center South, South and Sicilia) and control zone data for Germany (TenneT, TransnetBW, 50Hertz, Amprion), as data quality is better for these regions. For Denmark, we also retrieve bidding zone data, as one zone belongs to the Continental Europe area, while the other belongs to the Nordic area. 

Secondly, in the Continental Europe and the Nordic areas, we aggregate the region contributions until the final share of missing data points exceeds a threshold. We mark missing data points as "NaN" and set the sum of region contributions to NaN if at least one of them is NaN. Moreover, we set data points to NaN if they correspond to missing or corrupted measurements in the grid frequency time series, as these points cannot be used in our model. Thus, if the region contributions contain too many NaNs, we will not have many data points left in the aggregated data set. To avoid this problem, we sort the features and regions according to their share of NaN values and successively add up time series with an increasing share of NaNs. We stop the aggregation if the NaN share in the aggregated data set exceeds 30\% and do not add the remaining region contributions. For example, consider a (hypothetical) area consisting of 3 countries: A, B and C. For a given feature they have a share of 5\%, 30\% and 35\% NaNs. Then, we would compute the aggregated feature value by summing up the contributions of only A and B. As some entries are NaN in both A and B, we might only have a total of $\sim 30\%$ NaN in the aggregated data. 

\begin{figure*}
    \centering
    \includegraphics[width=\textwidth]{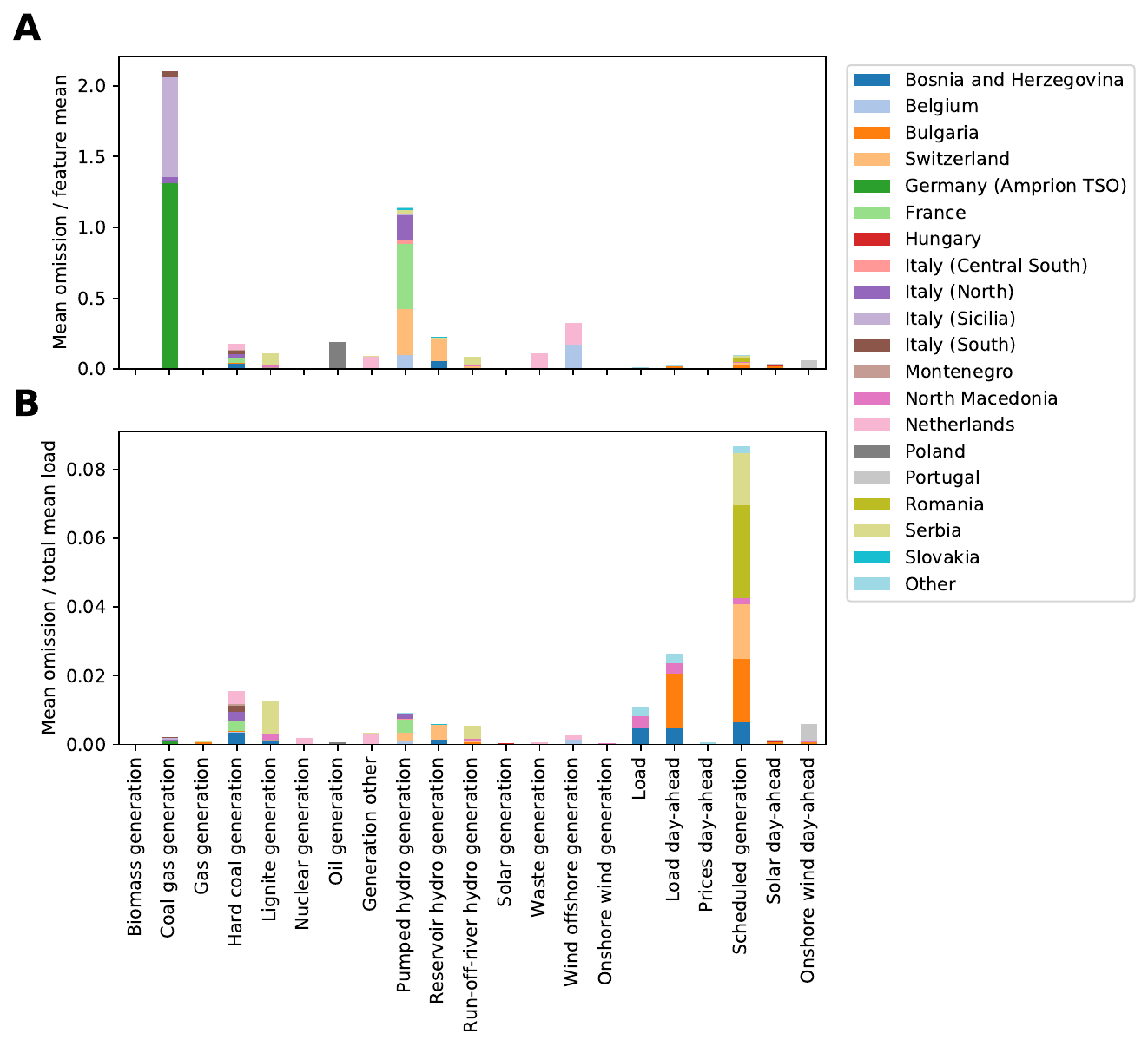}
    \caption{Omitted data contributions in Continental Europe. (a): We quantify the omitted values with the mean omitted feature value relative to the mean of the included features. The legend indicates the regions (mostly countries) where the omitted contributions come from. Regions with omitted (relative) contributions below 0.8\% are aggregated in the "Other" variable. (b) The mean omitted feature value relative to the total mean load of the synchronous area remains below 9\%.}
    \label{fig:supp_omitted_contributions}
\end{figure*}

This procedure omits a certain amount of data but allows us to retain a large sample size. In Continental Europe, most of the omitted contributions would increase the feature value by less than 50\% on average (Figure~\ref{fig:supp_omitted_contributions}A) and all omitted feature contributions are smaller than 9\% of the total mean load (Figure~\ref{fig:supp_omitted_contributions}B). In the Nordic area, we only omit the Finish day-ahead solar power forecast, accounting for 0.03\% of the area total mean load, and in GB there are no omissions. The aggregated data sets obtained from this procedure contain more than 29900 data points in each area (Table~\ref{tab:data_properties}). We thus generate large data sets to efficiently learn structures in the data, while still representing most of the load and generation within the areas. 

\begin{figure*}
    \centering
    \includegraphics[width=0.99\textwidth]{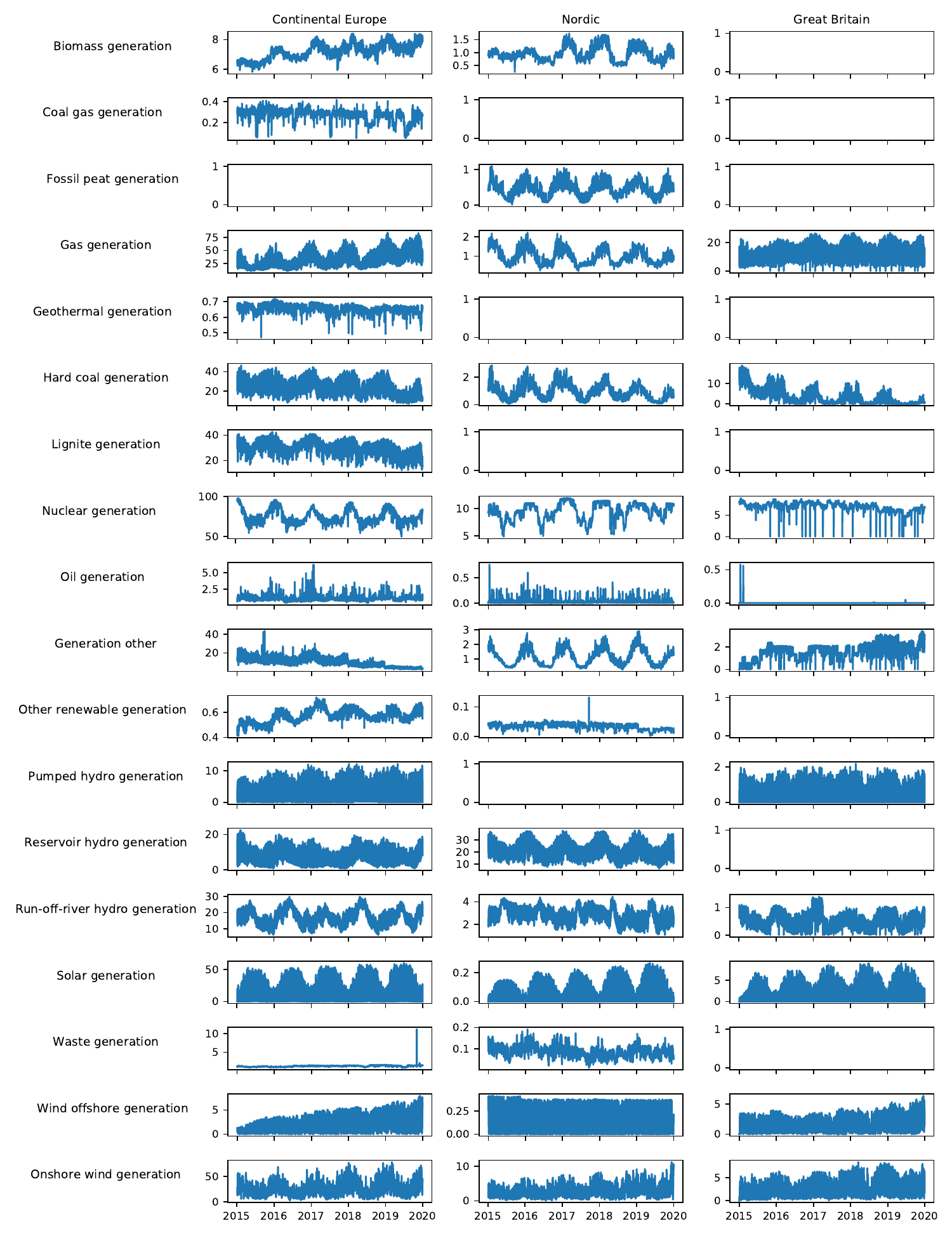}
    \caption{All (aggregated) time series from the ENTSO-E transparency platform \cite{ENTSOETransparencyPlatform}: Actual generation per type (in GW).}
    \label{fig:supp_gen_time_series}
\end{figure*}

\begin{figure*}
    \centering
    \includegraphics[width=\textwidth]{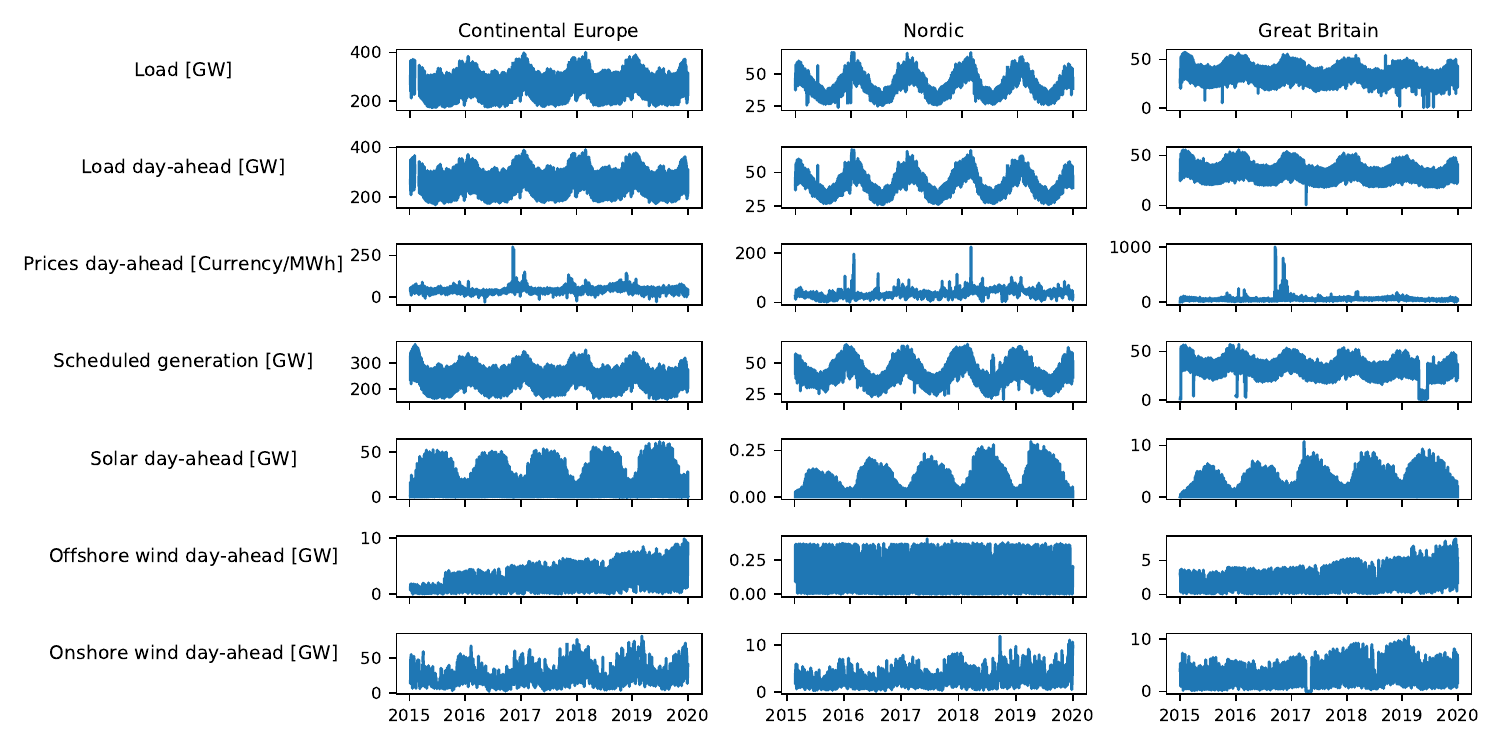}
    \caption{All (aggregated) time series from the ENTSO-E transparency platform \cite{ENTSOETransparencyPlatform}: All features except from actual generation per type.}
    \label{fig:supp_other_time_series}
\end{figure*}

Thirdly, we clean outliers in the (aggregated) features by visually inspecting their time series and removing isolated extreme values. For example, power generation from waste ("Waste generation") in Continental Europe exhibits a large outlier towards the end of 2019 (Figure~\ref{fig:supp_gen_time_series}) and Great Britain shows load values near zero in the same year (Figure~\ref{fig:supp_other_time_series}). Such outliers do not represent the typical energy system dynamics. They probably result from measurement errors and should be removed before fitting our Machine Learning model. The thresholds for removing outliers in the data set are defined in our python code \cite{frequency_xai_repo} and the cleansed data is available on zenodo \cite{kruseDataSetPredictionFrequency}.

Finally, we obtain 25 different (aggregated) times series of external features. Combining them with additional engineered features, such as forecast errors, we end up with 66 different external features (Table~\ref{tab:all_features}), which both contain day-ahead available features (such as the load forecast) and ex-post available features (such as the actual generation per type). None of the synchronous areas exhibits all 66 features and the number of model inputs thus varies between 50 and 64 (Table~\ref{tab:data_properties}).

\begin{table*}
    \small
    \centering
    \begin{tabularx}{\textwidth}{l| >{\raggedright}p{3cm} | X}
    \hline
        Ex-post & Ramps [MW/h] & Load ramp, Total generation ramp, Biomass ramp, Coal gas ramp, Fossil peat ramp,  Gas ramp, Geothermal ramp, Hard coal ramp, Lignite ramp,  Nuclear ramp, Offshore wind ramp, Onshore wind ramp, Oil ramp, Other ramp, Other renewables ramp, Pumped hydro ramp, Reservoir hydro ramp, Run-off-river hydro ramp, Solar ramp, Waste ramp  \\ \cline{2-3}
        {} & Generation and load [MW] & Load, Total generation, Synchronous generation, Biomass generation, Coal gas generation, Fossil peat generation, Gas generation, Geothermal generation, Hard coal generation, Lignite generation, Nuclear generation, Oil generation, Other generation, Other renewable generation, Pumped hydro generation, Reservoir hydro generation, Run-off-river hydro generation, Solar generation, Waste generation, Wind offshore generation, Wind onshore generation,  \\\cline{2-3}
        {} & Forecast errors of generation and load [MW] &  Forecast error load, Forecast error total generation, Forecast error solar,  Forecast error offshore wind,  Forecast error onshore wind \\\cline{2-3}
        {} & Forecast errors of ramps [MW/h] &  Forecast error load ramp, Forecast error generation ramp, Forecast error solar ramp, Forecast error offshore wind ramp, Forecast error onshore wind ramp \\\hline
        Day-ahead & Generation and load [MW] &  Load day-ahead, Scheduled generation, Solar day-ahead, Offshore wind day-ahead, Onshore wind day-ahead  \\\cline{2-3}
        {} & Ramps [MW/h] & Load ramp day-ahead, Generation ramp day-ahead, Solar ramp day-ahead, Offshore wind ramp day-ahead, Onshore wind ramp day-ahead \\\cline{2-3}
        {} & Other & Price ramp day-ahead [Currency/MWh/h], Prices day-ahead [Currency/MWh], Hour, Weekday, Month \\ \hline
    \end{tabularx}
    \caption{All external features in the data set. The units correspond to those used in our publicly available data set \cite{kruseDataSetPredictionFrequency}.}
    \label{tab:all_features}
\end{table*}

\begin{table*}
    \centering
    \begin{tabular}{c|c|c}
         Area & Number of features & Number of data points  \\ \hline 
         Continental Europe &  64 & 29913 \\
         Nordic & 58 & 37031 \\
         Great Britain & 50 & 41164 
    \end{tabular}
    \caption{Properties of our data sets}
    \label{tab:data_properties}
\end{table*}

\section*{Note S2: RoCoF extraction}
\label{sec:supp_note_rocof}
The Rate of Change of Frequency (RoCoF) is an indicator of frequency stability, which we use in our study. We extract the RoCoF at the beginning of each hour by smoothing the frequency increments with a rolling window of length $L$ and than looking for the maximum (absolute) RoCoF within a window of $\pm T$ around the full hour. 

\begin{figure*}
    \centering
    \includegraphics[width=\textwidth]{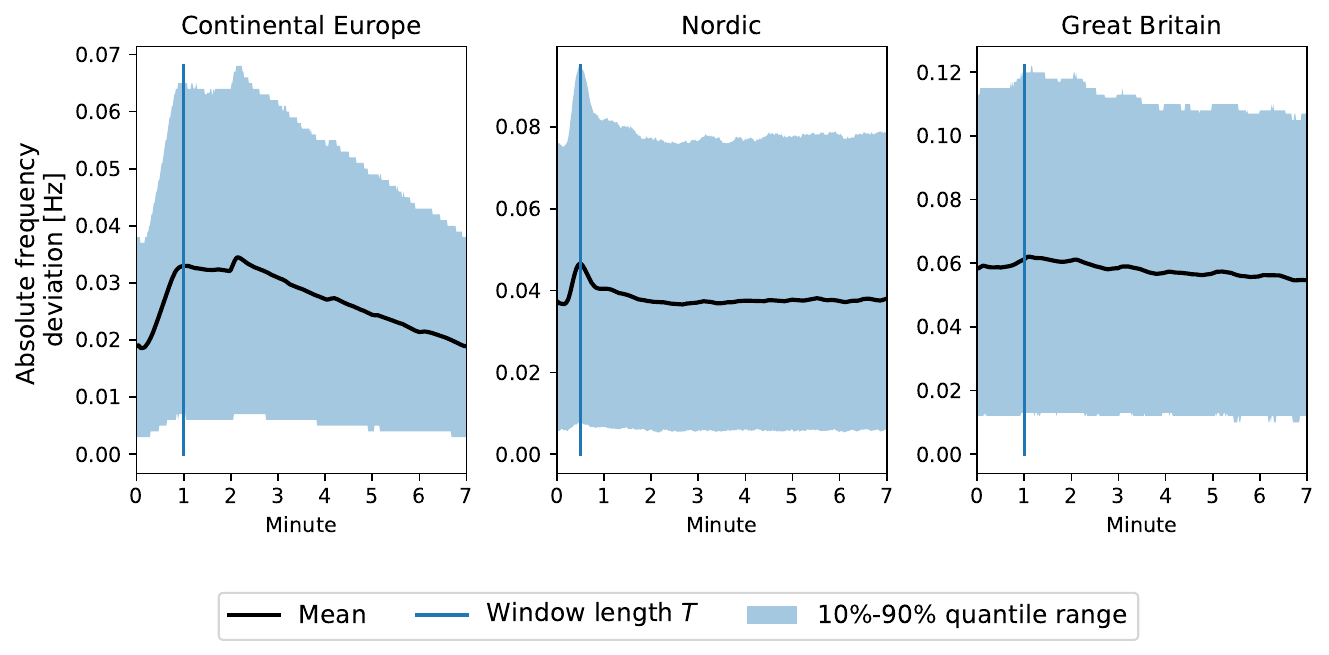}
    \caption{Evolution of hourly absolute frequency deviations. We display the mean evolution of the absolute frequency deviation during the first seven minutes of an hour. The deviation peaks at the beginning of the hour due to the impact of electricity trading \cite{weissbachHighFrequencyDeviations2009}. The time scale of this initial increase depends on the grid area. We choose the window length $T$ for the RoCoF extraction according to this time scale.}
    \label{fig:supp_rocof_time_scale}
\end{figure*}

We choose the values of $L$ and $T$ according to the typical time scale of the RoCoF in the three different synchronous areas. The average hourly evolution of the absolute frequency deviation indicates this time scale (Figure ~\ref{fig:supp_rocof_time_scale}). In Continental Europe and Great Britain, the average deviation reaches its maximum 60 s after the full hour, while the Nordic area exhibits its peak already after 30 s. We thus choose $L=T=60$ s in the Continental Europe and Great Britain areas, but a shorter time scale of $L=T=30$ s in the Nordic grid area. 

\section*{Note S3: Basic correlation analysis}
\label{sec:supp_note_corr_analysis}
A basic correlation analysis between external features and frequency stability indicators can already reveal interesting dependencies. However, this model-agnostic correlation analysis does not account for correlations among the features, which might affect the correlation coefficient between a feature and the stability indicator. Following the main text, we demonstrate this for the effect of nuclear ramps on the RoCoF in Continental Europe.

\begin{figure*}
    \centering
    \includegraphics[height=0.87\textheight]{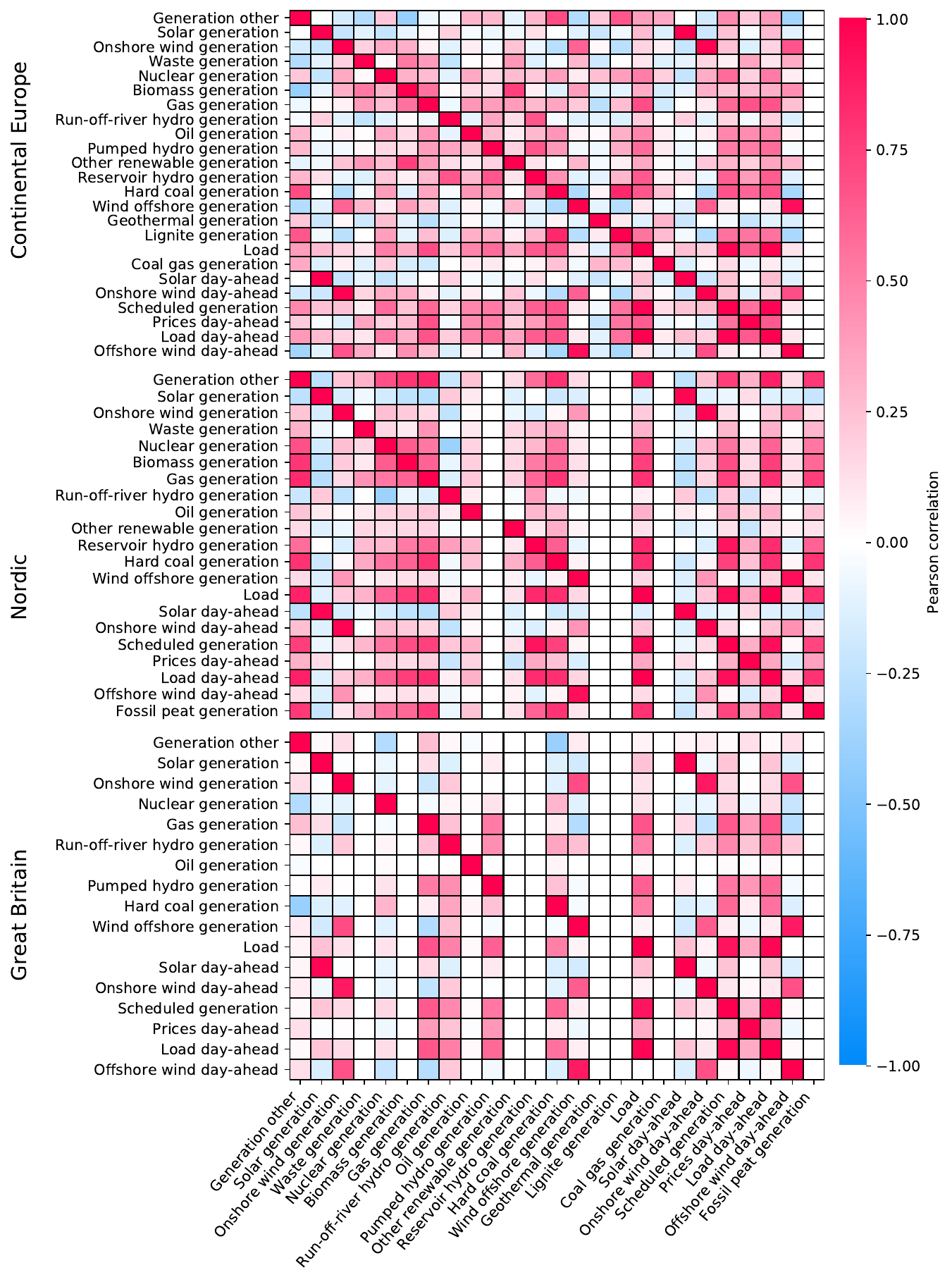}
    \caption{Pearson correlation coefficients between external features. To improve the visibility, we exclude our additional, engineered features from this plot. Features with 0 correlation everywhere have no data in a specific grid (e.g. Fossil peat data is only available in Nordic).}
    \label{fig:supp_feature_corr}
\end{figure*}

\begin{figure*}
    \centering
    \includegraphics[width=0.98\textwidth]{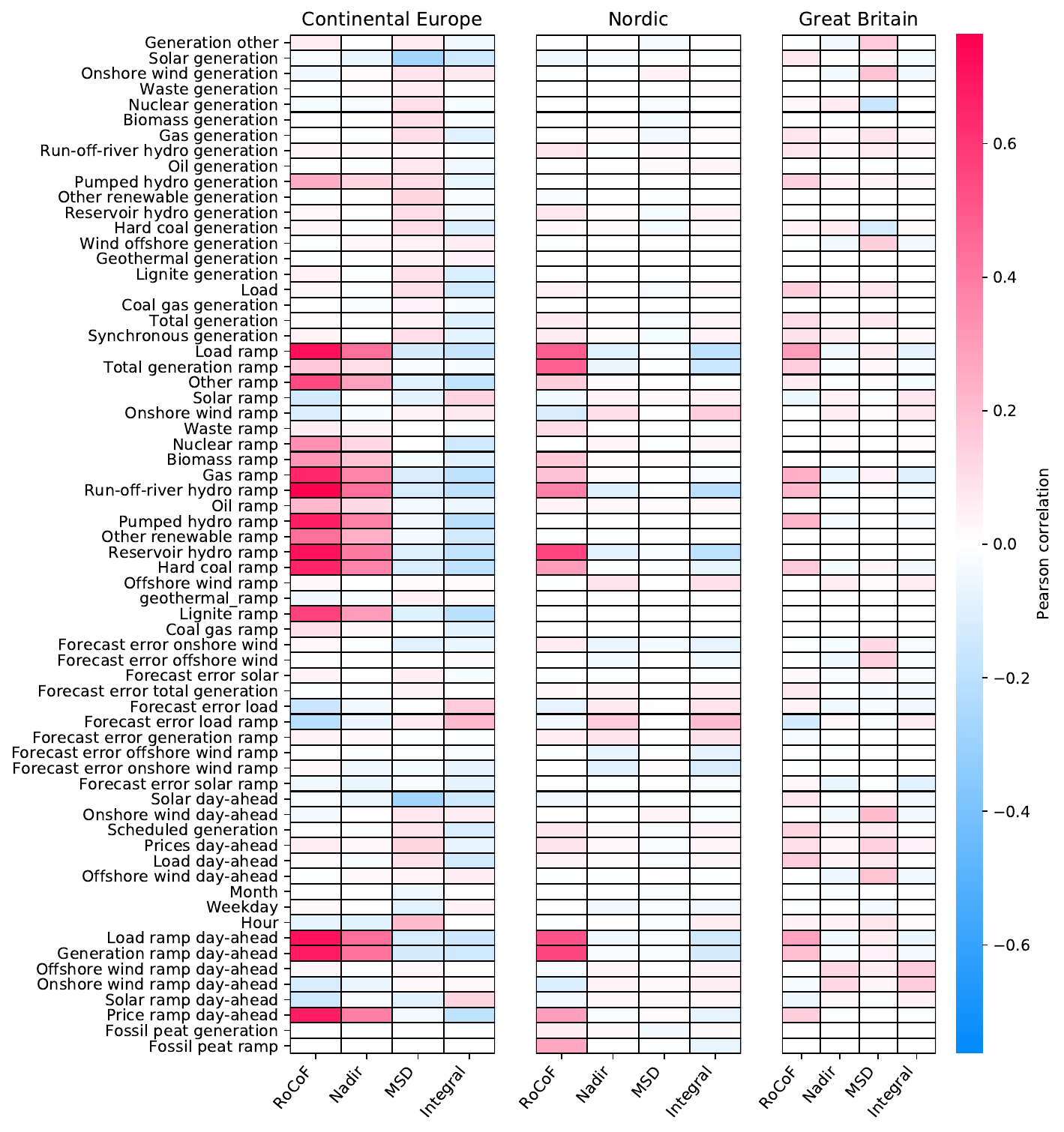}
    \caption{Pearson correlation coefficients between external features and frequency stability indicators. The scale of the colour code is adjusted to the maximum absolute correlation value $C_{max}$ and thus ends at $\pm C_{max}$.}
    \label{fig:supp_feature_target_corr}
\end{figure*}

As depicted in Figure~\ref{fig:supp_feature_corr}, there are various strong correlations between features in all three grid areas. For example, nuclear power generation is positively correlated with the load in Continental Europe. In Figure~\ref{fig:supp_feature_target_corr}, the features are correlated with our four stability indicators. We observe that nuclear ramps are positively correlated with the RoCoF in Continental Europe, which is not consistent with our SHAP results (see main text). We can explain the positive correlation of nuclear ramps with the hidden relationships to other variables, such as load ramps. Load ramps have a positive effect on the RoCoF in Continental Europe (Figure~\ref{fig:supp_RoCoF_dependencies}). Due to the strong correlation between load and nuclear power generation (Figure~\ref{fig:supp_feature_corr}) the effect of load ramps can thus "leak" into the correlation coefficients of nuclear ramps. This can explain why we observe a positive correlation between nuclear ramps and the RoCoF in Continental Europe, although nuclear ramps are RoCoF-offsetting in this area, as revealed by SHAP analysis in the main text. 

\section*{Note S4: Deterministic frequency deviations}
\label{sec:supp_note_dfds}
Deterministic frequency deviations (DFDs) occur at the beginning of electricity trading intervals \cite{weissbachHighFrequencyDeviations2009}. The generation is adapted in a step-wise manner at the beginning of these intervals, which are mostly hourly time periods. The mismatch between the step-wise generation and the continuously evolving load generates an instantaneous power imbalance, which causes a deterministic frequency jump at the beginning of the hour. 

Such DFDs are an important factor for frequency stability in Continental Europe. This is indicated by the time within the hour where the absolute frequency deviation peaks ("Nadir occurrence time"). Figure~\ref{fig:supp_nadir_dependencies} shows the histograms of these Nadir occurrence times within the hour. In Continental Europe, most of the Nadirs occur in the first five minutes, which indicates their strong connection to the deterministic electricity trading. In contrast, large deviations in Great Britain occur much more often during the hour and not only at the beginning. This indicates that DFDs play a smaller role in Great Britain than in Continental Europe. The Nordic area is in between, showing both strong hourly DFDs as well as Nadirs within the hour. 

\begin{figure*}
    \centering
    \includegraphics[width=\textwidth]{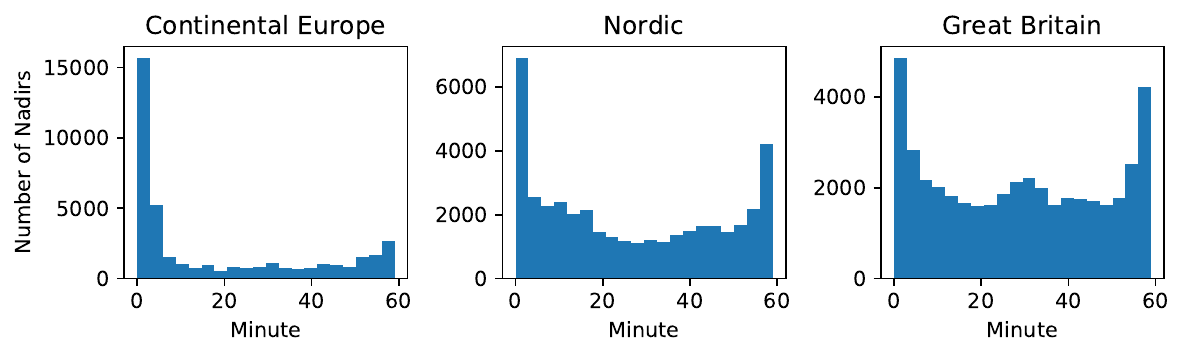}
    \caption{Distribution of Nadir occurrence time. The occurrence time of the nadir is the minute within the hour where the absolute frequency deviation reaches its peak. Its distribution within the hour varies between the grid areas, thus indicating the different importances of DFDs for the grid frequency dynamics. }
    \label{fig:nadir_minute}
\end{figure*}

\section*{Note S5: Performance evaluation of the machine learning model}
\label{sec:supp_note_performance}
We evaluate the performance of our Gradient Tree Boosting (GTB) model in terms of the R$^2$-score, which quantifies the proportion of variability explained by our model. A perfect prediction would result in a score of 1, while predicting the mean of the target results in a score of 0. As a benchmark, we compare the GTB model to the daily profile, which is an important null model for frequency dynamics. In particular, we quantify the gain over the daily profile, which is the model performance divided by the daily profile performance. Finally, we examine the importance of area-wide feature aggregation and the possibility to predict stability indicators day-ahead. Note that the GTB model used in our main text is referred to as the \textit{full model}. It builds on area-wide aggregated features containing both day-ahead \textit{and} ex-post available data. 

\begin{figure*}
    \centering
    \includegraphics[width=\textwidth]{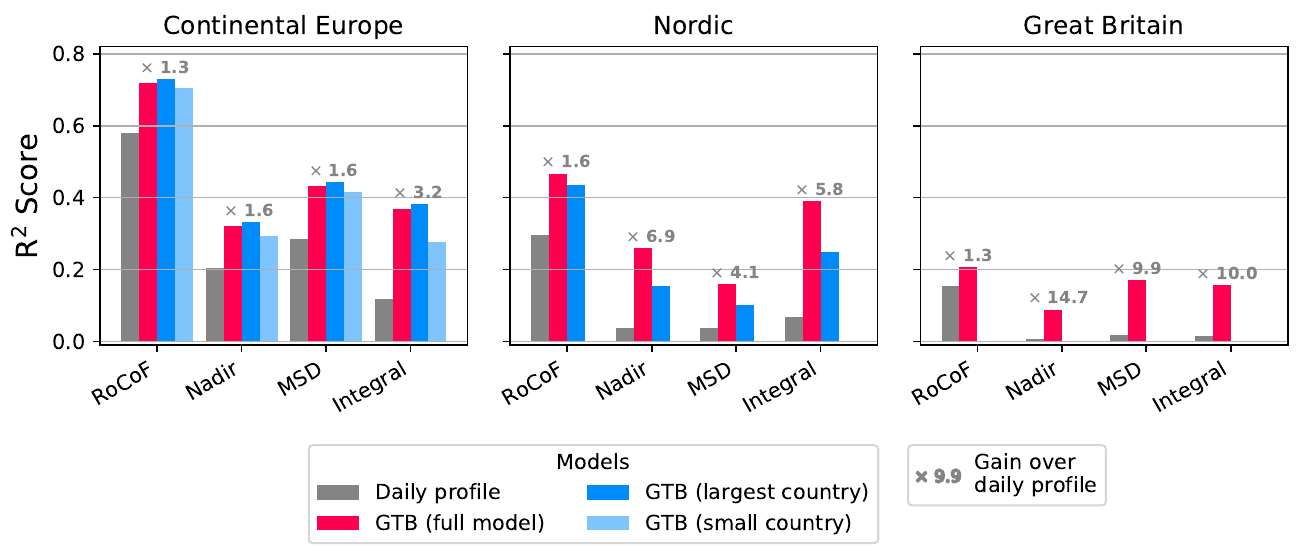}
    \caption{Performance of the full GTB model. The numbers on the bar plots indicate the gain of the GTB model over the daily profile. The full model comprises day-ahead and ex-post available features, which are aggregated area-wide. The country-level models use features from the largest country, i.e. the country with the largest power demand. In Continental Europe, we additionally introduce a model using data from a small country (with smaller average load), which is Switzerland. We retrieve the country-level data from the ENTSO-E transparency platform \cite{ENTSOETransparencyPlatform} and construct the same features as in the full model (Table~\ref{tab:all_features}). }
    \label{fig:supp_perf_vs_daily_profile}
\end{figure*}

The GTB model performs best in Continental Europe, while the performance gain over the daily profile is largest in Great Britain. Figure~\ref{fig:supp_perf_vs_daily_profile} displays the R$^2$-score for each stability indicator and each area. We obtain the best predictions in Continental Europe ($R^2 \sim 0.7$) and the lowest scores in Great Britain. In contrast, the performance gain over the daily profile is largest in Great Britain (maximum 14.7) and smallest in Continental Europe (maximum 3.2), while the Nordic area is in between (maximum 6.9).  Frequency dynamics in Continental Europe are rather deterministic compared to the stochastic dynamics in Great Britain. Therefore, the prediction is easier and the additional gain through Machine Learning is smaller in Continental Europe. Consistently, the GTB performance is best for the RoCoF as this indicator most strongly reflects the hourly deterministic frequency jumps. 

The model performance depends on whether we choose area-wide aggregated features or country-level data (Figure~\ref{fig:supp_perf_vs_daily_profile}). In the Nordic area, we obtain a lower performance when using data from only the largest country (Sweden) instead of aggregating it area-wide. As the grid frequency is affected by all locations within the grid, it is not surprising that data aggregation is important. In Continental Europe, the largest country model (using Germany) performs equally well (even a bit better) than the aggregated model, but choosing a smaller country (Switzerland) also reduces performance. The frequency stability in Continental Europe is particularly affected by large load and generation ramps (see main text). Therefore, already a country responsible for the largest ramps can yield good performance. Overall, the area-wide feature aggregation yields better or similar results than regional data among the areas.  

\begin{figure*}
    \centering
    \includegraphics[width=\textwidth]{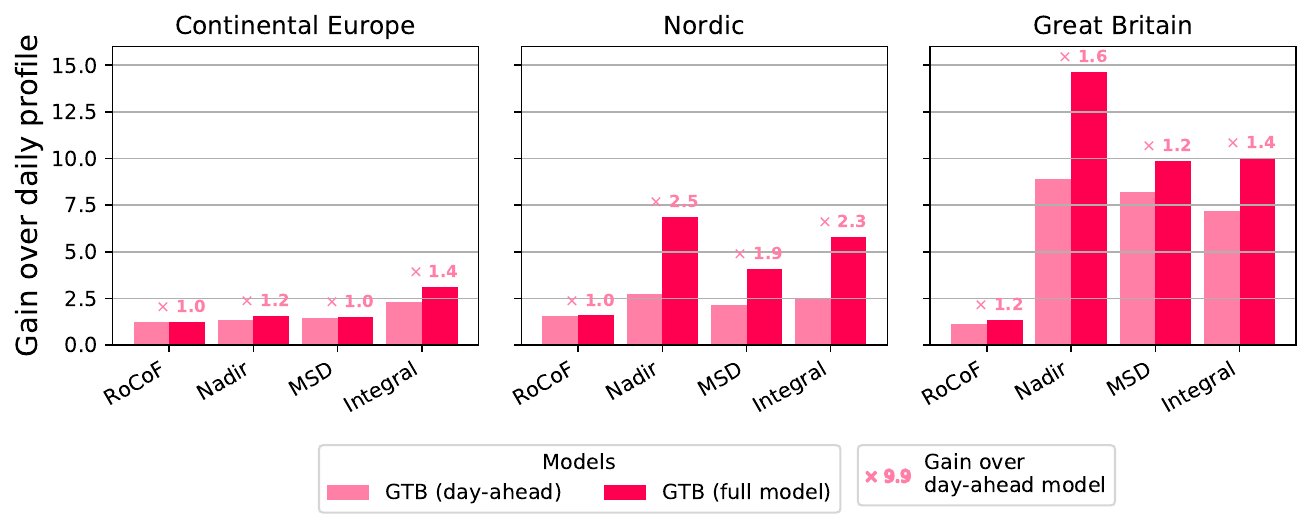}
    \caption{Performance of the day-ahead GTB model. We compare the full model and the day-ahead model in terms of their gain over the daily profile. The numbers on the bar plots indicate the gain of the full model over the day-ahead model. }
    \label{fig:supp_perf_vs_post_hoc}
\end{figure*}

Using only day-ahead available data in our GTB model already outperforms the daily profile for all stability indicators and areas (Figure~\ref{fig:supp_perf_vs_post_hoc}). In Great Britain, the day-ahead model exhibits the strongest performance gain over the daily profile (maximum 8.9), followed by the Nordic area (maximum 2.8) and Continental Europe (maximum 2.3). However, adding ex-post data in the full model can strongly improve the performance, especially in the Nordic area. We quantify this effect in terms of the gain over the day-ahead model, i.e. the full model performance divided by the day-ahead model performance. In the Nordic area, the gain of the full model over the day-ahead model is the largest (maximum 2.5), while it is lowest in Continental Europe (maximum 1.4). The benefits of adding ex-post data in the Nordic area stems from the importance of forecasting errors for the prediction (see main text). 


\section*{Note S6: Additional results with SHAP values}
\label{sec:supp_note_shap_results}
\begin{figure*}
	\centering
	\includegraphics[height=0.9\textheight]{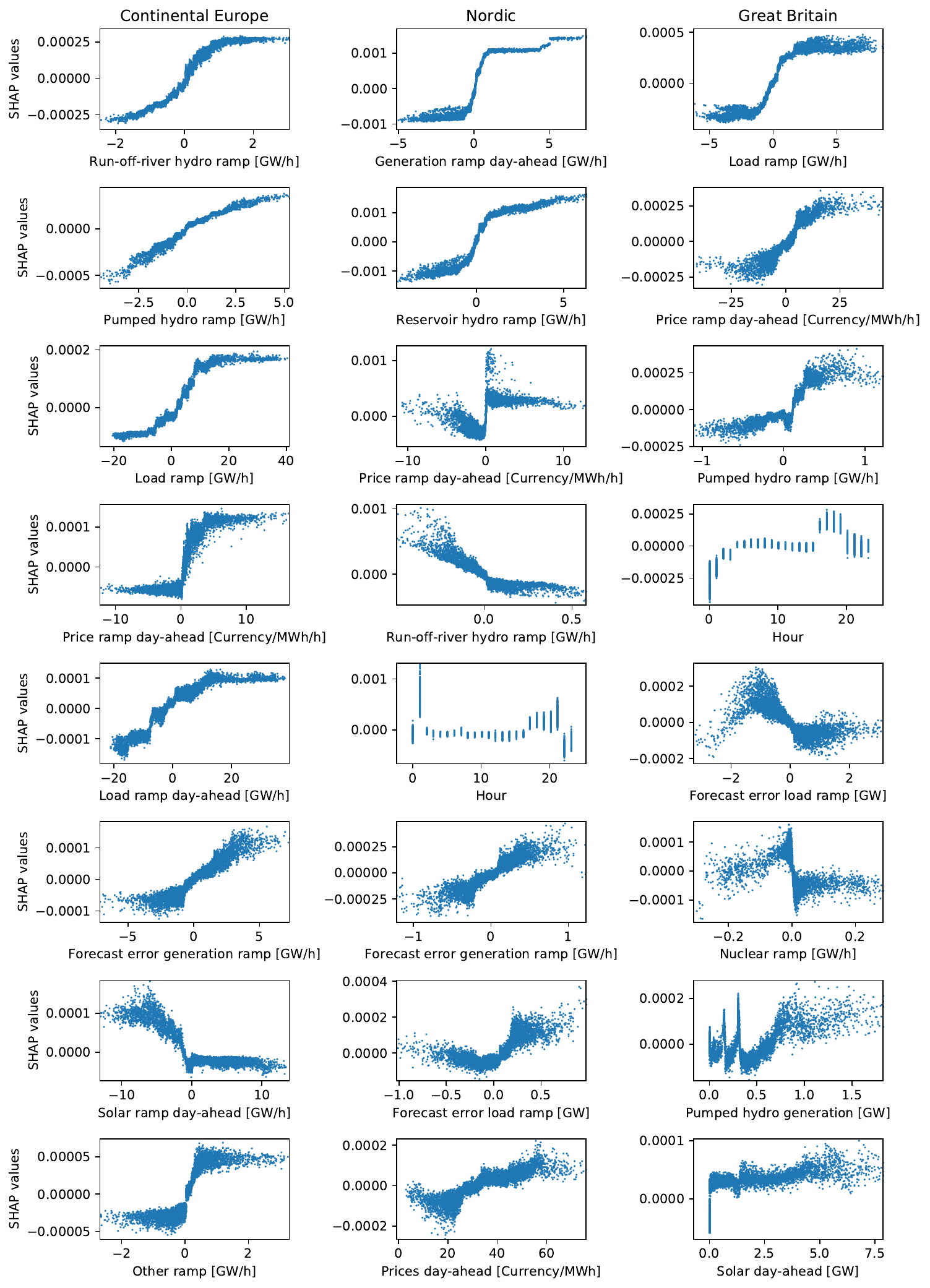}
	\caption{RoCoF dependency plots for the eight most important features.}
	\label{fig:supp_RoCoF_dependencies}
\end{figure*}

\begin{figure*}
	\centering
	\includegraphics[height=0.9\textheight]{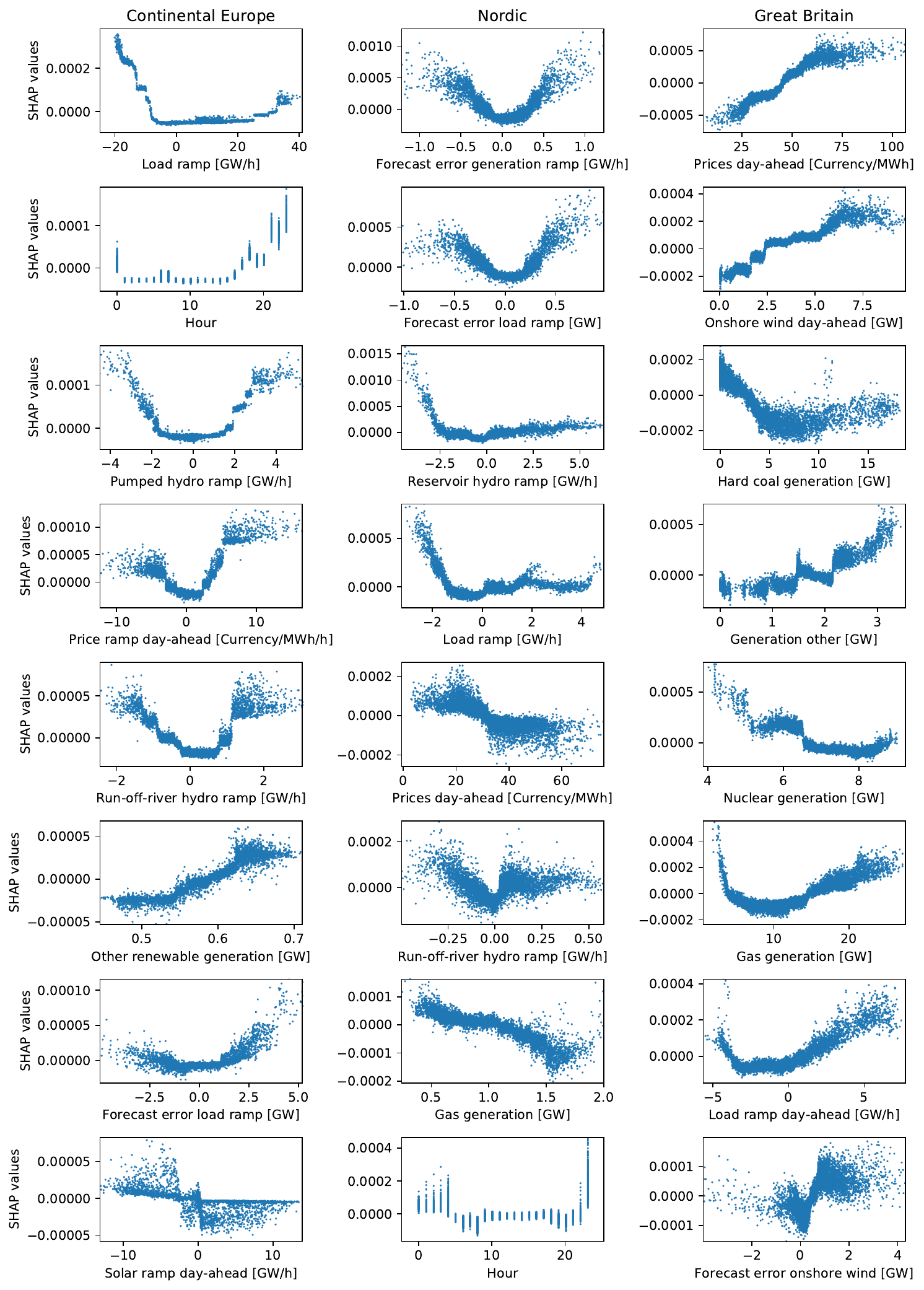}
	\caption{MSD dependency plots for the eight most important features.}
	\label{fig:supp_msd_dependencies}
\end{figure*}

\begin{figure*}
	\centering
	\includegraphics[height=0.9\textheight]{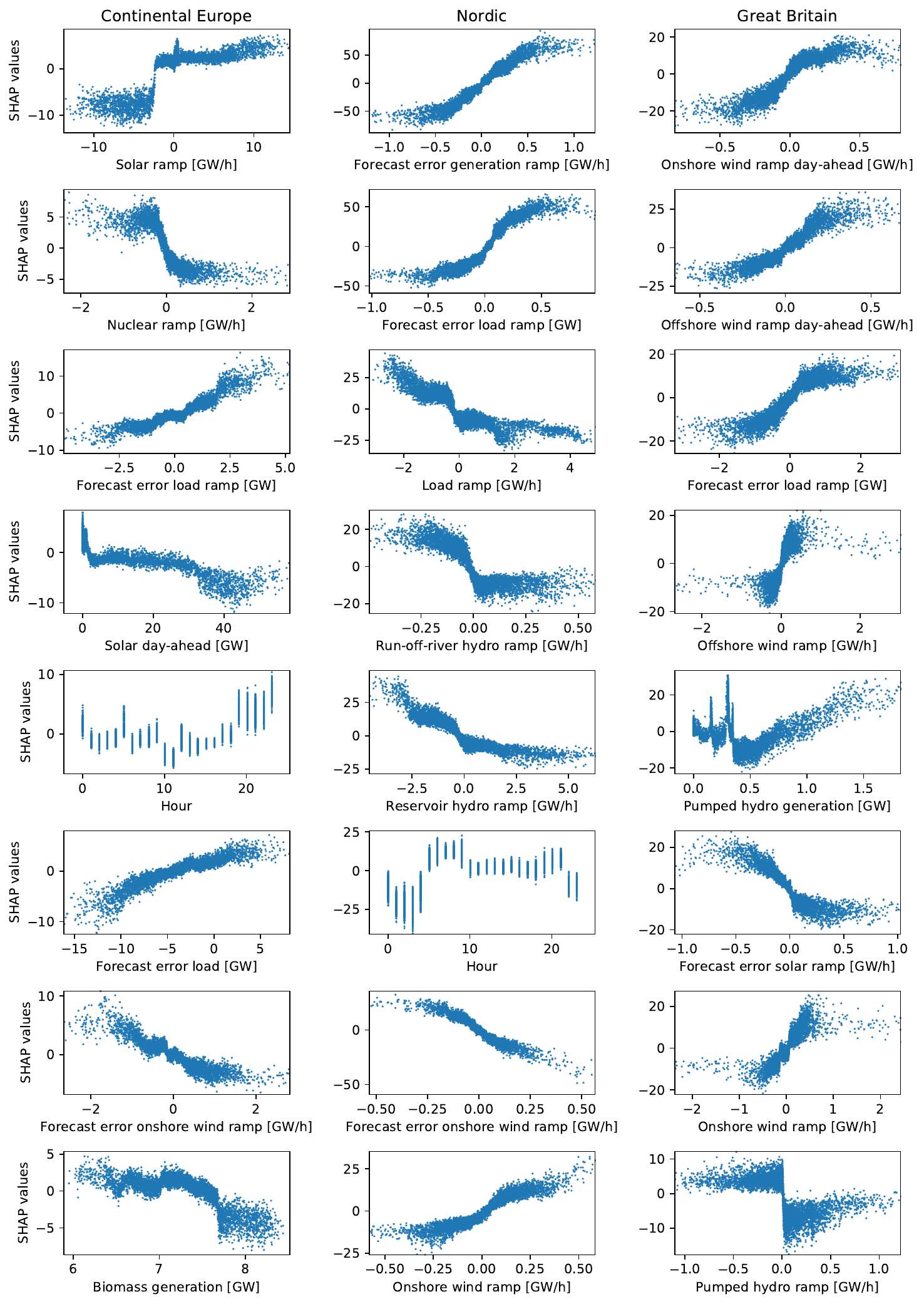}
	\caption{Nadir dependency plots for the eight most important features.}
	\label{fig:supp_nadir_dependencies}
\end{figure*}

\begin{figure*}
	\centering
	\includegraphics[height=0.9\textheight]{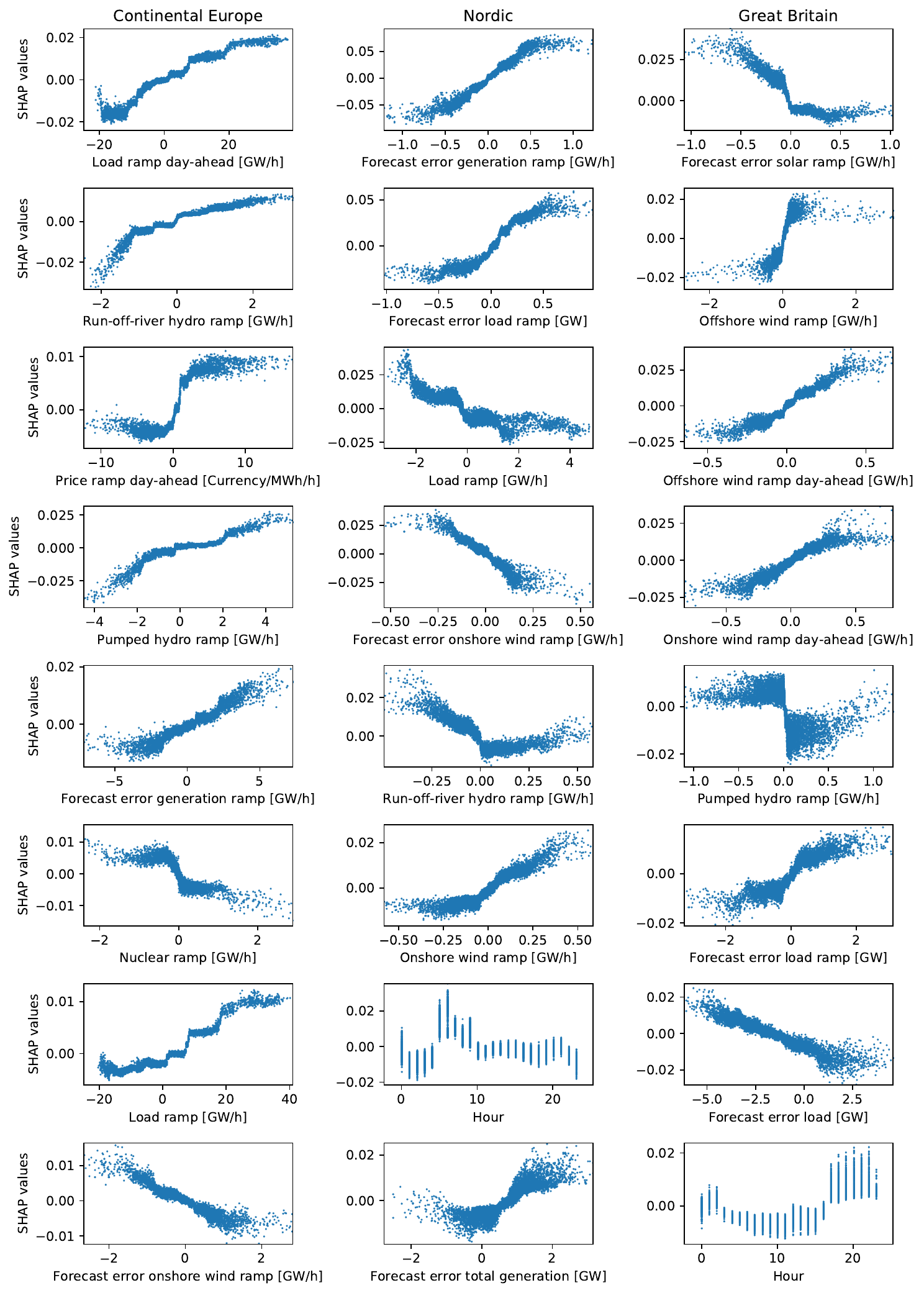}
	\caption{Integral dependency plots for the eight most important features.}
	\label{fig:supp_integral_dependencies}
\end{figure*}

We use SHAP values to explain our Machine Learning model for frequency stability indicators. An overview of the most important SHAP dependencies in our model is available in Figures \ref{fig:supp_RoCoF_dependencies}, \ref{fig:supp_msd_dependencies}, \ref{fig:supp_nadir_dependencies} and \ref{fig:supp_integral_dependencies} for each of the four stability indicators. In many cases, we observe non-linear dependencies, which underlines the importance of using a non-linear, complex Machine Learning model such as Gradient Tree Boosting.

\begin{figure*}
	\centering
	\includegraphics[width=\textwidth]{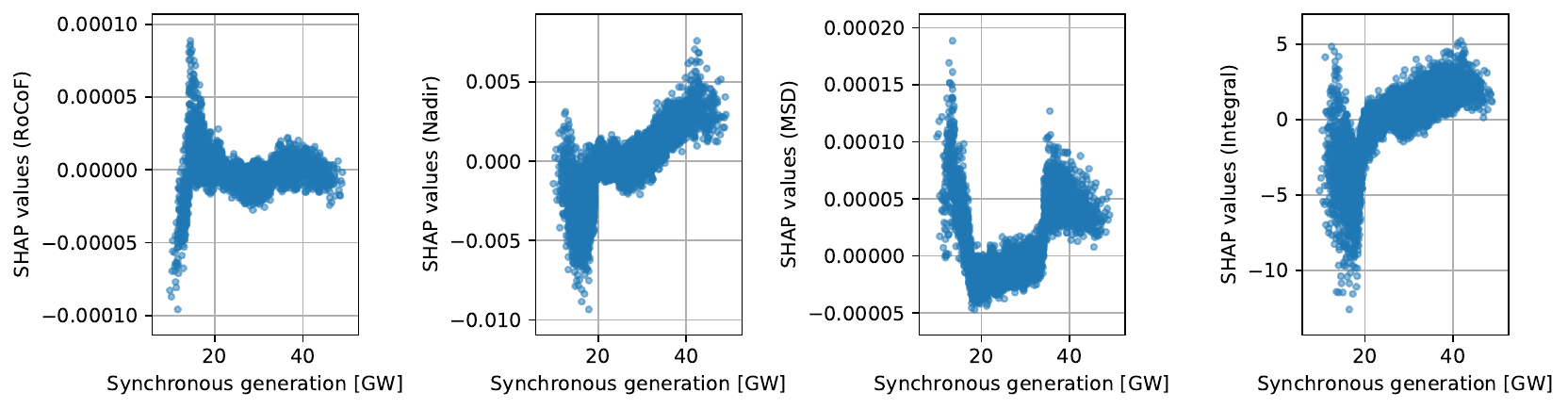}
	\caption{Dependency plots of synchronous generation in Great Britain. Each subplot corresponds to one of our aggregated indicators for frequency stability.}
	\label{fig:supp_gb_inertia}
\end{figure*}

Here, we further discuss the effect of synchronous generation on frequency stability indicators. The (total) synchronous generation, which we use as a proxy for the total inertia within the power grid, is not among the eight most important features (Figures \ref{fig:supp_RoCoF_dependencies}, \ref{fig:supp_msd_dependencies}, \ref{fig:supp_nadir_dependencies} and \ref{fig:supp_integral_dependencies}). The relative feature importance of the synchronous generation remains below 0.09 in Continental Europe, below 0.06 in the Nordic area and below 0.20 in Great Britain. Overall, the average effect of the (approximated) inertia on the aggregated stability indicators is thus relatively low compared to the most important features. Among the areas, the total synchronous generation is most important in Great Britain, which is consistent with the high share of renewable energy sources in the British power system and the resulting low-inertia situations \cite{milanoFoundationsChallengesLowInertia2018}. The effect of the inertia in Great Britain is depicted in the dependency plots of Figure~\ref{fig:supp_gb_inertia}. For all stability indicators, we observe the maximum (absolute) effect of the synchronous generation at low feature values. This is particularly evident for the RoCoF, where the effect of values larger than 20 GW is zero on average. In conclusion, the (approximated) inertia mostly affects frequency stability in Great Britain in extreme situations of low inertia, but the average effect of this feature on our aggregated stability indicators is negligible. 

\clearpage

\end{document}